\documentclass{aa}
\usepackage{graphicx}
\usepackage{hyperref}
\usepackage{txfonts}
\usepackage{xcolor}
\usepackage{caption}
\usepackage{subcaption}
\usepackage{textcomp}
\usepackage[rightcaption]{sidecap}
\usepackage[mathlines]{linenoaa}
\usepackage{placeins}  
\usepackage{media9}

\newcommand{\CellWithForceBreak}[2][c]{
\begin{tabular}[#1]{@{}c@{}}#2\end{tabular}}

\begin{document} 
    \title{DEM analysis of the 6 September 2011 large-scale coronal wave}
    \titlerunning{DEM analysis of the 6 September 2011 large-scale coronal wave}	
	
    \author{Amaia Razquin
          \inst{1}
          \and
          Astrid M. Veronig\inst{1, 2}
          \and
          Karin Dissauer\inst{1, 3}
    }

    \institute{University of Graz, Institute of Physics, Universitätsplatz 5, 8010 Graz, Austria \\
        \email{amaia.razquin-lizarraga@uni-graz.at}
        \and 
            University of Graz, Kanzelh\"ohe Observatory for Solar and Environmental Research, Kanzelh\"ohe 19, 9521 Treffen, Austria
        \and
             NorthWest Research Associates, 3380 Mitchell Lane, Boulder, CO 80301, USA
    }
	
    \date{Received March, 2026; accepted June, 2026}
	
    \abstract 
    {Large-scale coronal waves are globally propagating intensity enhancements in extreme-ultraviolet (EUV) and soft X-ray (SXR) observations, in association with solar flares and coronal mass ejections (CMEs). They are interpreted as low-coronal signatures of a large-amplitude fast magnetosonic wave. On 6 September 2011, a fast ($v\approx 1000$ km~s$^{-1}$) large-scale coronal wave accompanied an eruptive X2.1 class flare. A notable feature of this event was the temporary disappearance of a segment of the wave front in EUV channels sensitive to quiet-Sun plasma, while the same structure remained visible in higher temperature channels. 
    }
    {We analyse the plasma properties associated with this large-scale coronal wave and quantify its impact on the local plasma as it passed through the corona. We aim to determine whether the temperature increase at the wave front can be explained solely by a compressive wave, and how the observed spatial variability relates to the plasma temperature distribution.
    }  
    {We evaluate plasma parameters at multiple locations along the wave front and across its propagation path. We apply differential emission measure (DEM) diagnostics to SDO/AIA EUV observations to derive local density, temperature, emission measure (EM), and DEM distributions, and examine their temporal evolution during the wave passage. 
    }
    {The wave passage causes increases of 6-8\% in density and 10-18\% in temperature. While the density increase is comparable to earlier reports, the temperature increase exceeds expectations. This indicates that the temperature enhancement cannot be explained by compressional adiabatic heating alone, and instead suggests the presence of additional heating mechanisms, such as magnetic reconnection or wave mode conversion. During the temporary disappearance of the wave, the plasma parameters at the wave front increase, but with a strong spatial variability, with density increases as low as 1\% and as high as 10\%. The DEM distribution shows that the initial temperature in the affected area is notably higher than typical quiet-Sun regions ($\bar{T}\gtrsim1.7$~MK), which allows plasma to be heated beyond the peak response of the AIA 193 and 211~\AA~channels. We conclude that the apparent temporary disappearance of the wave front is primarily due to the combined effects in the intensity of the CME-associated coronal dimming following the wave and the wave itself, with heating further reducing its detectability in channels sensitive to quiet-Sun temperatures.
    }
    {}
	
    \authorrunning{A. Razquin et al.}
    \maketitle 
    \nolinenumbers

\section{Introduction} \label{sec:introduction} 
Large-scale coronal waves are globally propagating wavelike disturbances in the solar corona that are observed as enhancements in the extreme-ultraviolet (EUV) and soft X-ray (SXR) emission of the Sun. They arise from the impulsive energy release during solar eruptions, such as flares and coronal mass ejections (CMEs) \citep{warmuth2007large}. 
 
The physical nature of large-scale coronal waves has been the subject of a long debate. The first observations of large-scale coronal waves were made by the Extreme-ultraviolet Imaging Telescope (EIT, \citet{delaboudiniere1995eit}) on board the Solar and Heliospheric Observatory (SOHO, \citet{domingo1995soho}) \citep{moses1997eit, thompson1998soho}. 
Initially, they were interpreted as the coronal counterparts of chromospheric Moreton waves \citep{moreton1960recent}, which are observed as arc-shape fronts propagating away from a flaring active region (AR). In this wave model, large-scale coronal waves are interpreted as fast-mode magnetosonic waves \citep{uchida1968propagating, thompson1999soho, wang2000eit, warmuth2001evolution, wu2001three}. However, discrepancies in the speeds \citep{klassen2000catalogue} and spatial extent \citep{eto2002relation} between large-scale coronal waves and Moreton waves, as well as the existence of stationary bright fronts \citep{delanee1999cme}, prompted the development of non-wave models. In these models, large-scale coronal waves are interpreted as a result of the successive restructuring of the magnetic field lines during a CME eruption \citep{delanee1999cme, chen2002evidence, attrill2007coronal}. 

Unifying these interpretations, \citet{chen2002evidence} proposed a hybrid interpretation consisting of both a fast-mode wave and a slower non-wave component caused by the continuous stretching and restructuring of magnetic fields, which was later coined hybrid model \citep{nitta2013large}. Subsequent statistical studies \citep{warmuth2011kimnematical} provided evidence of multiple types of large-scale coronal waves,  and event studies showed the coexistence of these two components \citep{chen2011first, asai2012first, kumar2013multiwavelength}. Overall, the current picture suggests that large-scale coronal waves may be composed of both a fast-mode magnetosonic wave and a non-wave component, although the non-wave component can also be interpreted to be related to the expanding CME bubble rather than the wave pulse itself. A detailed discussion on the different models of large-scale coronal waves and their arguments can be found in recent reviews \citep{zhukov2004nature, willsdavey2009eit, gallagher2011large, zhukov2011eit, patsourakos2012nature, liu2014advances, warmuth2015large, chen2016global, long2017understanding, zheng2024recent}.

The formation of large-scale coronal waves is generally understood to result from the sudden expansion of closed magnetic loops during a CME eruption. The wave front is observed to propagate ahead of the laterally expanding CME flanks, which act as the primary driver of the wave’s motion \citep{keinreich2009stereo, patsourakos2009what, veronig2008high, ma2011observations, downs2012understanding}. After the initial impulsive phase, the CME flanks decelerate, allowing the large-scale coronal wave to continue propagating freely at speeds close to the fast magnetosonic speed in the corona \citep{veronig2008high, patsourakos2009what, warmuth2011kimnematical, cheng2012investigation, mann2023propagation}. 

Large-scale coronal waves are often accompanied by type II radio bursts \citep{mann1999coronal, klassen2000catalogue, nitta2014relation, muhr2014statistical}, which indicate the presence of a shock wave in the corona. Behind the bright wave front, sometimes regions of reduced intensity are observed. These low-intensity areas correspond either to rarefaction regions that formed behind the compressive wave \citep{muhr2011analysis, liu2019impacts} or to coronal dimmings \citep{zhukov2004nature, podladchikova2005automated, dissauer2018statistics}. Coronal dimmings are observed when plasma is evacuated as a consequence of the eruption of a magnetic flux rope and the subsequent expansion of the overlying magnetic fields, leading to a sudden drop in coronal emission \citep{hudson1996long, sterling1997yohkoh, thompson1998soho}; for a review see \citet{veronig2025coronal}. Although the dimmings may be primarily related to the main eruption in the AR, the large-scale coronal waves can also induce reconfiguration that opens field lines and generates localised dimming signatures \citep{zhou2020magnetic, arazquin2026selected}. This interpretation also aligns with non-wave models, whereas the formation of rarefaction regions is well understood as a consequence of a compressive fast-mode wave \citep{warmuth2015large}, consistent with the expansion and upwards plasma flows that are observed behind large-scale coronal wave fronts \citep{harra2011spectroscopic, veronig2011plasma}. 

Large-scale coronal waves are known to interact with and influence the local plasma and magnetic structures they propagate through. Waves get reflected \citep{gopalswamy2009euv, kumar2013eruption}, refracted \citep{willsdavey1999observations, shen2012evidence} and transmitted \citep{olmedo2012secondary, shen2013diffraction} at regions of increased magnetosonic speed such as the boundaries of ARs and coronal holes. 
Other interactions can manifest as oscillations and deflections of filaments \citep{okamoto2004filament, zhang2024transverse}, coronal loops \citep{shen2014chain, qu2017observations}, streamers \citep{tripathi2007on, li2025observations}, and cavities \citep{liu2012quasi}, as well as the triggering of jets \citep{shen2014chain} and sympathetic eruptions \citep{schrijver2013pathways}. 

Typical large-scale coronal wave speeds range from 200–600~km~s$^{-1}$ \citep{klassen2000catalogue, warmuth2011kimnematical, muhr2014statistical}, although some events can reach velocities exceeding 1000~km~s$^{-1}$. They are generally observed to heights up to about 50–100 Mm \citep{keinreich2009stereo, patsourakos2009what, delannee2014time, podladchikova2019three}, and they are most commonly observed in EUV passbands corresponding to plasma temperatures of approximately 1.4–2~MK. Large-scale coronal waves were first directly observed in the 195~\AA~channel of SOHO/EIT, dominated by the Fe XII line with a peak response at about 1.4~MK. Subsequent detections were performed in 171~\AA~(Fe IX; $T\approx0.8$~MK) and 284~\AA~(Fe XV; $T\approx2$~MK). The seven EUV channels of the Atmospheric Imaging Assembly (AIA; \citet{Lemen2012aia}) onboard the Solar Dynamics Observatory (SDO; \citet{pesnell2012sdo}) allow detailed thermal characterisation, spanning temperatures from 0.1~MK to 10~MK. In AIA observations, large-scale coronal waves are most clearly seen in the 193~\AA~and 211~\AA~channels, which correspond to plasma at 1.6–2~MK, while they appear progressively weaker from 335 to 94, 131, and 304~\AA. It has also been observed that waves appear as a decrease in intensity in 171~\AA~images \citep{willsdavey1999observations, long2011wave, schrijver2011february}. This anticorrelation between 171~\AA~and 193 or 211~\AA~intensities arises because the passage of the wave heats the coronal plasma, shifting the differential emission measure (DEM) toward higher temperatures. As a result, channels sensitive to hotter plasma (193 and 211~\AA) record an enhanced emission, whereas cooler channels (171~\AA) show a reduced or negative response. This effect provides clear evidence that large-scale coronal waves involve both compression and plasma heating of the plasma environment passed by the wave. 

The first successful spectroscopic analysis of a large-scale coronal wave was conducted by \citet{asai2008strongly}, using the EUV Imaging Spectrometer (EIS) onboard Hinode, who reported blueshifts of 100~km~s$^{-1}$ and heating of plasma to temperatures above 2~MK.  
\citet{veronig2011plasma} carried out a dedicated spectroscopic observing campaign in a sit-and-stare mode and were able to obtain plasma densities and flows at the wave front of a large-scale coronal wave. They reported downflows of 20~km~s$^{-1}$ and upflows of 5~km~s$^{-1}$, interpreted as plasma compression and relaxation of the plasma over the wave passage. The associated density compression was estimated to be below 10\%. The same event was also studied in \citet{harra2011spectroscopic} who reported a wave propagation speed of about 500 km~s$^{-1}$ along the EIS slit.

With the advent of SDO, the broad spectral coverage of AIA has been utilised to better characterise plasma compression and heating during large-scale coronal wave events. \citet{schrijver2011february} employed AIA observations alongside their response functions and showed that the anticorrelation between the different intensity evolution in the channels was consistent with adiabatic heating of the plasma. \citet{kozarev2011offlimb} were the first to apply DEM analysis with AIA data to an EUV wave, deriving a lower limit for plasma compression of 12--18\%. Similarly, \citet{vanninathan2015coronal} investigated two EUV waves and found density increases of 6--9\% and temperature rises of 5--6\%, consistent with adiabatic heating. Their DEM analysis further explained the dark fronts observed in 171~\AA~images as a consequence of plasma heating from the temperature range near the instrument’s peak response to values with a lower response, supporting the results of \citet{schrijver2011february}. 

In this study, we analysed the large-scale coronal wave associated with the X2.1 flare on 6 September 2011 using DEM analysis. We examined the coronal plasma response to the passage of the large-scale coronal wave and investigated the spatial variation along different segments of the wave front. This spatially resolved approach allowed us to capture the fine-scale variability of the wave, including regions where the wave temporarily disappears in certain EUV channels, and to quantify differences in plasma density, temperature, and emission measure across different sectors. By comparing the observed temperature increases with the predictions of adiabatic compressional heating, we assessed whether compression alone can account for the thermal response of the coronal plasma. Furthermore, we examined how the instrumental response functions of different AIA channels determine the observed intensity variations, providing insight into why the wave appears more or less visible in specific wavelengths.

\section{Event overview and analysis} \label{sec:observations}
\subsection{Event overview} \label{sec:event}
On 6 September 2011, a globally propagating large-scale coronal wave was observed in association with an X2.1 flare and a CME event originating from AR~11283 (NOAA coordinates N14~W18). The GOES soft X-ray flare started at 22:12~UT and reached its peak intensity around 22:21~UT \citep{janvier2016evolution}. A halo CME was subsequently detected with a speed of approximately 990~km~s$^{-1}$ \citep{feng2013magnetic, dissauer2016projection}. The eruption was accompanied by the development of a complex coronal dimming \citep{dissauer2018detection, vanninathan2018plasma, prasad2020mhd, veronig2025coronal}.

The large-scale coronal wave was first analysed by \citet{nitta2013large} and determined to have a maximum speed of $v\approx 1250$~km~s$^{-1}$. A later study by \citet{dissauer2016projection} detected speeds closer to $1070$~km~s$^{-1}$ in the northward direction. Notably, \citet{dissauer2016projection} identified the disappearance of a segment of the EUV wave front at 22:21:48~UT which subsequently reappeared at approximately 22:22:48~UT. This disappearance is observed in the AIA 171, 193, and 211~\AA~channels, while the wave remains clearly visible in the hotter 335~\AA~channel. A Type II radio burst was detected during the event \citep{nitta2014relation}. Despite the strength of the eruption and the wave amplitude, no associated Moreton wave was observed \citep{dissauer2016projection}. 

\subsection{Data and data reduction} \label{sec:data}

To analyse the large-scale coronal wave, we used the data from the six coronal EUV filters (94, 131, 171, 193, 211, 335~\AA) of SDO/AIA. These filters cover a broad temperature range, effectively being sensitive to plasma between $10^5$--$10^7$~K \citep{Lemen2012aia, boerner2012initial}. 

We processed the AIA data following the procedure described by \citet{purkhart2023multipoint}. The EUV images were deconvolved using the SolarSoft IDL (SSWIDL) function \texttt{aia\_deconvolve\_richardsonlucy} to mitigate the effects of stray light during a flare. The point spread function for each wavelength channel was obtained using \texttt{aia\_calc\_psf}, and all deconvolutions were performed with the default parameter settings. This step enhances image contrast and sharpness, thereby improving the sharpness and contrast of the reconstructed DEM curves and the potential of the subsequent plasma diagnostics. The deconvolved images were then processed with standard SSWIDL routines (\texttt{aia\_prep}).

For the reconstruction, we prepared input map structures containing the images from the six EUV channels. Each image was differentially rotated to a pre-event time, chosen as 22:12:07~UT. In order to increase the signal-to-noise ratio, we rebinned each image by $4\times4$ pixels, resulting in an effective spatial resolution of 2.35~arcsec per pixel for the output DEM. Since we analysed quiet-Sun regions, we used the regular exposure images and discarded the lower exposure ones taken when AIA was in flare mode. This results in an effective time cadence of 24~s in the DEM reconstruction. Notably, the different channels of AIA do not take images exactly cotemporally. Thus, when combining the six channels to reconstruct a single DEM, there is a maximum lag of 8~s between the first and the last image.

\subsection{Differential emission measure analysis} \label{sec:dem}
The DEM is the distribution of emitting plasma as a function of temperature along the line-of-sight (LOS) $h$. It is an inherent property of the plasma arising from the photon-producing collisions of free electrons with target photons and heavier ions \citep{brian2024emission}. Assuming that the coronal plasma is optically thin and in thermal equilibrium, the DEM is defined as
\begin{equation}
    \text{DEM}(T)=n(h(T))^2\frac{\mathrm{d}h}{\mathrm{d}T},
\end{equation}
where $n$ is the electron density. For an instrument observing at a wavelength $\lambda$, the observed intensity $I_\lambda$ can be described as
\begin{equation}
I_\lambda=\int_{T}{R_\lambda(T)\text{DEM}(T)\mathrm{d}T},
\end{equation}
where $R_\lambda(T)$ is the temperature response function of the instrument. Thus, the measured counts in each pixel are a convolution of the DEM with the instrument response. The reconstruction of the DEM from observations is an ill-posed inverse problem, and multiple algorithms have been developed to achieve it.

\citet{hannah2012differential} developed a regularised inversion method to reconstruct the DEM from multi-channel observations. We used the IDL implementation of this method to derive the DEM from six AIA EUV channels sensitive to coronal temperatures (94, 131, 171, 193, 211, 335~\AA). For the reconstruction, we set the temperature range from 0.1 to 10~MK and divided it into 50 evenly spaced logarithmic temperature bins. We used the \texttt{aia\_get\_response} procedure to calculate the AIA temperature response function at the event time using coronal abundances from the CHIANTI 10.0 database \citep{dere1997chianti, zanna2021chianti}.

Once we derived the DEM, we calculated the total emission measure (EM) over all temperature bins ($\Delta T$) as
\begin{equation}\label{eq:em}
    \text{EM}=\sum_{T_0}^{T_1}{\text{DEM}(T)\cdot\Delta T},
\end{equation}
following the definition by \citet{cheng2012differential}. Similarly, we calculated the emission weighted mean temperature $\bar{T}$ and the plasma density $\rho$ with the following expressions:
\begin{equation}\label{eq:t}
    \bar{T}=\frac{\sum_{T_0}^{T_1}{\text{DEM}(T)\cdot T \cdot \Delta T}}{\text{EM}}
\end{equation}
and
\begin{equation}\label{eq:n}
    \rho=\sqrt{\frac{\text{EM}}{h}},
\end{equation}
where for an estimate of the contributing length scale $h$ we used the hydrostatic scale height of coronal plasma (e.g. \citet{aschwanden2005physics}):
\begin{equation}
    h=\frac{2k_B\bar{T}}{\mu m_H g},
\end{equation}
where $k_B$ is the Boltzman constant, $\mu$ is the mean molecular weight, $g$ is the gravitational acceleration at the photosphere, and $m_H$ is the hydrogen atom mass.
We assumed fully ionised plasma with $\mu=0.64$, $m_H=1.67\times10^{-27}$~kg, and $g=274$~m~s$^{-1}$, which resulted in $h\approx170$ Mm (with slight variations depending on the location). 

\begin{figure*}[t!]
    \centering
    \resizebox{\hsize}{!}{\includegraphics{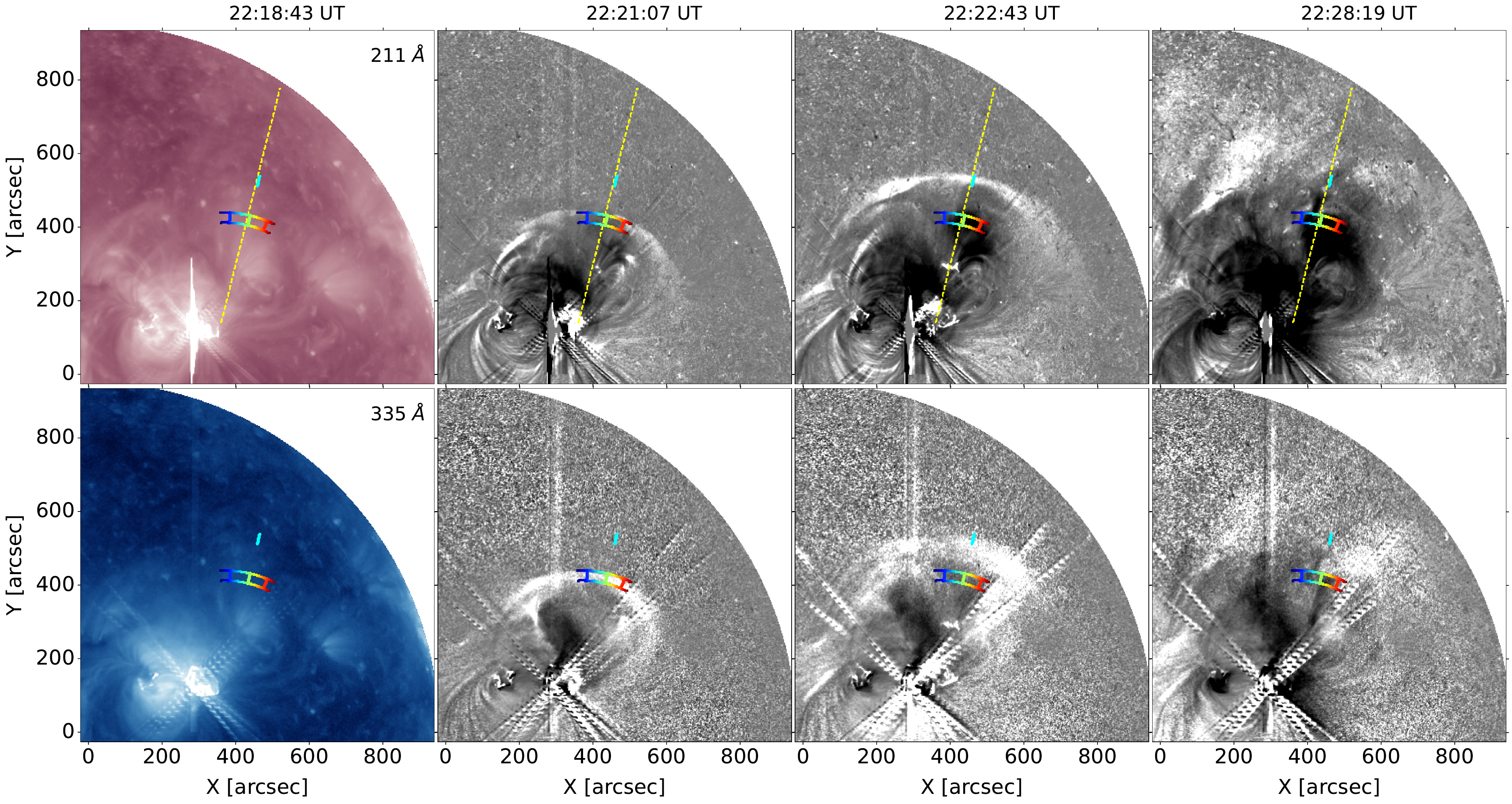}}
        \caption{Overview of the EUV wave event on 6 September 2011 observed in SDO/AIA 211 and 335~\AA~base-ratio images. The rainbow-colour contours outline the arc-shaped ROI defined to encompass the wave front at 22:21:07~UT. Three representative patches within the ROI are marked in red (west), green (centre), and blue (east). An additional patch selected further along the propagation path at 22:22:43~UT, is shown in cyan (north). In the 211~\AA~observations (top row) a yellow line marks the slice where the intensity for Fig.~\ref{fig:aia_overview}b is calculated. The associated movie is available online.}
        \label{fig:overview}
\end{figure*}

Since large-scale coronal waves are known to be formed within the thermal range of about 1--2.5~MK and we were looking into quiet coronal regions, we calculated the plasma parameters (EM, $\bar{T}$, and $\rho$) from the extracted DEM in the temperature bins between 0.3 and 6.8~MK, that is, $T_0=0.3$~MK and $T_1=6.8$~MK in Eqs.~(\ref{eq:em}--\ref{eq:n}). From the derived DEM maps, we also calculated background-subtracted DEM ($\Delta$DEM) maps by subtracting a pre-event DEM to isolate the contribution of the wave.

\section{Results} \label{sec:results} 

Figure~\ref{fig:overview} and the accompanying animation show the evolution of the large-scale coronal wave on 6 September 2011 in the 211 and 335~\AA~channels of AIA alongside the patches where we investigated the coronal response. An extended version of this figure and animation showing all AIA wavelengths used for the DEM reconstruction is given in Fig.~\ref{fig:app:overview}. We calculated the plasma parameters in four main patches parallel to the propagating wave front. In addition, to analyse the wave disappearance and the spatial variability of the coronal response we selected a region of interest (ROI), delineated by the rainbow contours in Fig.~\ref{fig:overview}, within which we extracted the plasma parameters in evenly distributed segments.   

\subsection{Regions of interest and EUV observations} 
\begin{figure}
    \centering
    \resizebox{\hsize}{!}{\includegraphics{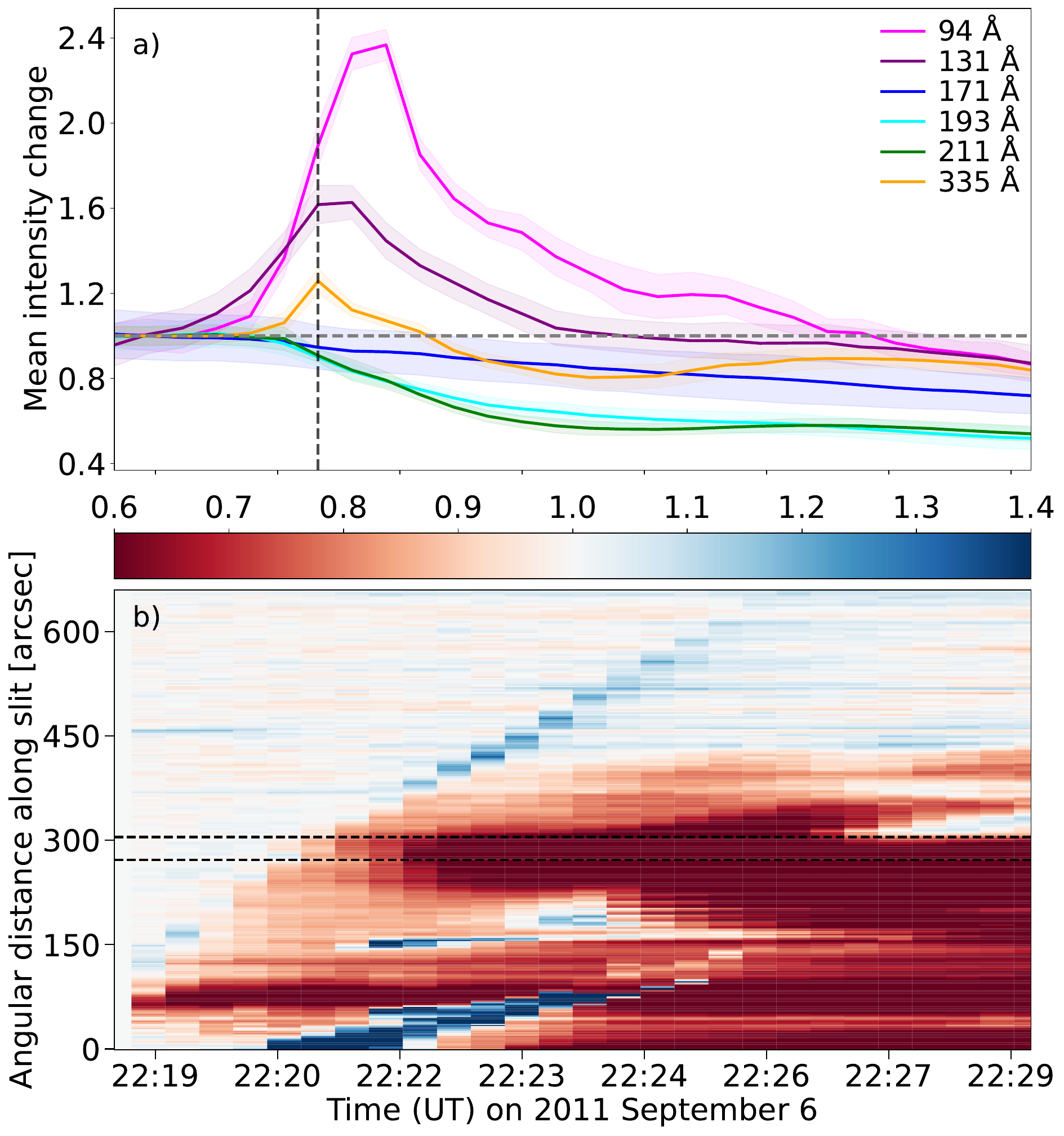}}
    \caption{(a) Evolution of the mean intensity within the ROI for all coronal AIA EUV filters. Shaded areas denote the standard deviation of the intensity, and the vertical dashed line marks the time of the wave passage at 22:21:07~UT. (b) Distance-time plot of the base-ratio AIA 211~\AA~intensity along the slit indicated by the yellow line in Fig.~\ref{fig:overview}. Two horizontal black lines mark the portion of the slit corresponding to the ROI.}
    \label{fig:aia_overview}
\end{figure}

We performed the analysis of the large-scale coronal wave in two sets of regions. On the one hand, we considered four patches where we analysed in detail the change in plasma parameters (see Table~\ref{tab:parameters}) and the DEM curve over the event time frame. On the other hand, we selected an extended ROI covering a large fraction of the wave front in order to study the spatial variability of the plasma parameters and the wave disappearance.

The extended ROI is an arc-shaped ROI perpendicular to the wave trajectory.
The ROI's radial extent is indicated by the rainbow-colour contours in Fig.~\ref{fig:overview}. We positioned the ROI over a region where the wave is visible in the 335~\AA~channel but not in the 211~\AA~channel to additionally investigate the nature of the temporary wave disappearance. We therefore defined the ROI using the AIA observations at 22:21:07~UT as reference. We limited the length of the arc to avoid contamination from diffraction patterns and detector blooming (see the movies accompanying Figs.~\ref{fig:overview} and \ref{fig:app:overview}).

The large-scale coronal wave observed on 6 September 2011 propagated at speeds exceeding 1000~km~s$^{-1}$. This speed is sufficiently fast for spatial variations in the wave position to be observed within the multi-wavelength set of images used for the DEM reconstruction, which have a maximum temporal separation of 8~s  (see Sect.~\ref{sec:data}). To ensure that the ROI encompasses the wave passage simultaneously in all images that are combined, we set the arc width to 20~Mm in the direction of propagation. 

Figure~\ref{fig:aia_overview}a presents the temporal evolution of the mean intensity within the ROI for all coronal AIA EUV filters over the duration of the event. The shaded areas indicate the corresponding standard deviations, while the vertical dashed line marks the passage of the wave at 22:21:07~UT. The mean intensity in the ROI increases by approximately 20\% in 335~\AA, 60\% in 131~\AA, and 140\% in 94~\AA, whereas a decrease is observed in the 193 and 211~\AA~channels. A decrease is also present in 171~\AA, although it is more gradual and affected by higher noise levels. We note that, while the wave front is visible in the 131 and 94~\AA~channels (see Fig.~\ref{fig:app:overview}) and thus part of the increase observed in Fig.~\ref{fig:aia_overview}a is indeed due to the wave, a large contribution to the increase is caused by further flaring in the AR (see animation accompanying Fig.~\ref{fig:app:overview}). The intensity variations do not occur simultaneously across all wavelengths, which is caused by the offset between frames of different wavelengths. This effect is particularly evident in the delayed peak observed in 94~\AA~and the early decrease in 193~\AA. 

The decrease in mean intensity observed in the 193 and 211~\AA~channels within the ROI is primarily associated with the abrupt disappearance of the wave and does not represent a general property of large-scale coronal wave propagation. This behaviour is illustrated in Fig.~\ref{fig:aia_overview}b, which shows the temporal evolution of the base-ratio intensity along the long slit indicated by the yellow dashed line in Fig.~\ref{fig:overview}, as observed by the AIA 211~\AA~channel. Before 22:20~UT, the wave is visible as an enhancement in intensity (shown in blue). Subsequently, the intensity increase weakens and eventually becomes a decrease in emission, before the wave reappears with a stronger enhancement, particularly after 22:22~UT. Following the wave passage, the intensity decreases largely due to the coronal dimming that develops after the eruption and is expanding away from the eruption centre. The two horizontal black lines in Fig.~\ref{fig:aia_overview}b delimit the portion of the slit corresponding to the ROI. Within this interval, the wave cannot be identified, and the observed evolution is dominated by the coronal dimming. Figure~\ref{fig:app:aia_overview} shows the same distance-time intensity plot for all AIA EUV wavelengths and illustrates that the coronal dimming dominates the evolution in the channels sensitive to quiet-Sun temperatures (171, 193, 211~\AA). Thus, the decrease in the AIA 193 and 211~\AA~channels in Fig.~\ref{fig:aia_overview}a is caused by the coronal dimming that develops at higher levels of the corona and overtakes the wave propagation, producing a net decrease in emission. The increase in intensity observed in in the 131 and 94~\AA~channels in Fig.~\ref{fig:aia_overview}a, caused by further flaring in the AR, can also be clearly distinguished in Fig.~\ref{fig:app:aia_overview}. 

We performed a detailed analysis on the plasma parameters and DEM curve evolution on four representative patches shown in red, green, blue and cyan in Fig.~\ref{fig:overview}. The patches are placed perpendicular to the wave front and have a width of 20 Mm to encompass the full width of the wave front. Three of these patches lie within the ROI previously defined and are shown in red, green and blue.
We selected these patches to sample a region where, in the AIA 211~\AA~channel at 22:21:07~UT, the large-scale coronal wave is clearly visible (red patch), a region where the wave is still visible but less pronounced (blue patch), and a region where the wave is absent (green). This approach allows us to compare the coronal response to the large-scale coronal wave simultaneously at different locations along the front, and better understand the nature of its temporary disappearance. For comparison, we also selected a patch further in the wave propagation, at 22:22:43~UT, where the wave is clearly visible in both the AIA 211 and 335~\AA~channels. This patch is shown in cyan in Fig.~\ref{fig:overview}. Hereafter, the red, green, and blue patches are referred to as west, centre and east patches due to their relative location within the ROI, and the cyan patch is referred to as the north patch. The arc-shaped region is referred to as the ROI.

\begin{figure*}[t!]
    \centering
    \resizebox{\hsize}{!}{\includegraphics{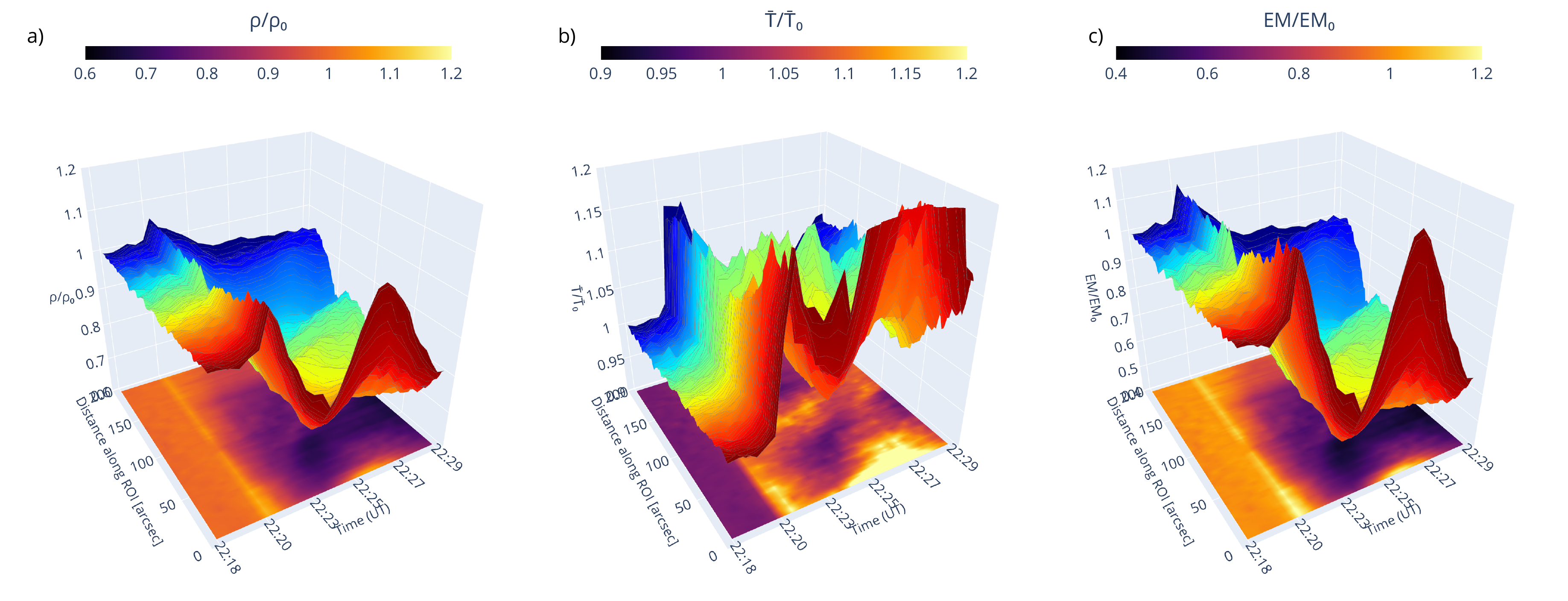}}
        \caption{Relative evolution of the plasma parameters derived from the DEM analysis within the arc-shaped ROI. We show the temporal variation of (a) density $\rho$, (b) mean weighted temperature $\bar{T}$, and (c) emission measure EM for 100 segments across the wave front propagation direction along the ROI. Individual curves are colour-coded by position along the arc, from west (red) to east (blue), consistent with the contour colour scale in Fig.~\ref{fig:overview}. All quantities are divided by their pre-event values, computed as the average over the first five time steps (22:18–22:19~UT). The lower portion of each panel displays a distance–time heatmap of the corresponding relative changes. The associated movie is available online.}
        \label{fig:plasma_parameters}
\end{figure*}

\subsection{Plasma parameters and their evolution} \label{sec:res:param}

The DEM analysis reveals a clear but highly non-uniform plasma response to the large-scale coronal wave along the arc-shaped ROI. All three parameters -- density, mean weighted temperature, and emission measure -- show a distinct peak at the wave passage, but with strong spatial variability: the western sectors exhibit the largest increases, while the central sectors show localised dips and prolonged temperature enhancements. While density and emission measure decrease soon after the wave has passed, temperature shows a stronger and more sustained increase, in all locations exceeding adiabatic predictions. 

\begin{table*}
    \caption{\label{tab:parameters}Plasma parameters at time of EUV wave passage for selected patches}   
    \centering
    \begin{tabular}{cccccccccc}         
    \hline
    Time [UT] & patch & Location & \CellWithForceBreak{$\rho$ \\ ($10^{8}$~cm$^{-3}$)} & $\rho/\rho_0$&  \CellWithForceBreak{$\bar{T}$ \\ ($10^6$~K)} &  $\bar{T}/\bar{T}_0$  &  \CellWithForceBreak{EM \\ ($10^{26}$~cm$^{-5}$)} &  $\text{EM}/\text{EM}_0$ & \CellWithForceBreak{$\Delta$DEM (min, max) \\ ($10^{20}$~cm$^{-3}$K$^{-1}$)} \\
    \hline
     22:21:07 & West & N31W35 & $1.76 \pm 0.11$ & $1.08$ & $2.16 \pm 0.15$ & $1.18$ & $5.4 \pm 0.6$ & $1.16$ & ($1.12$, $-0.94$) \\
     22:21:07 & Centre & N32W33 & $1.78 \pm 0.10$ & $1.01$ & $2.21 \pm 0.08$ & $1.15$ & $5.8 \pm 0.6$ & $1.02$ & ($0.41$, $-0.93$) \\
     22:21:07 & East & N33W29 & $2.05 \pm 0.10$ & $1.06$ & $1.98 \pm 0.11$ & $1.13$ & $6.9 \pm 0.7$ & $1.12$ & ($0.51$, $-0.38$) \\
     22:22:43 & North & N39W39 & $1.71 \pm 0.05$ & $1.08$ & $1.64 \pm 0.04$ & $1.10$ & $4.1 \pm 0.2$ & $1.16$ & ($0.71$, $-0.24$) \\
    \hline 
    \end{tabular}
    \tablefoot{We list the time, patch and location where the plasma parameters are calculated, and the corresponding density $\rho$, density increase $\rho/\rho_0$, mean weighted temperature $\bar{T}$, temperature increase $\bar{T}/\bar{T_0}$, emission measure EM, emission measure increase $EM/EM_0$, and maximum and minimum DEM change $\Delta\text{DEM}$.}
\end{table*}

Figure~\ref{fig:plasma_parameters} shows the relative evolution of the (a) density $\rho$, (b) mean emission weighted temperature $\bar{T}$, and (c) emission measure EM in the ROI extracted using the DEM analysis as described in Sect.~\ref{sec:dem}. We obtained the plasma parameters for 100 segments parallel to the wave propagation along the ROI, and colour-coded them from blue to red based on their position in the arc as shown in the contours in Fig.~\ref{fig:overview}; that is, red lines show the plasma parameters in the westernmost sector of the ROI, blue lines show the parameters for the easternmost sector, while green and yellow lines are roughly in the centre sectors of the ROI. We used the average of the first five time steps (22:18 to 22:19~UT) to calculate the pre-event parameter values and calculated the relative change of all time steps by dividing each value by the pre-event level. At the bottom boundary of each plot we show the relative change of each parameter as a distance-time heatmap, where the distance reflects the position of the segments along the arc. This distance-time heatmap is also directly shown in Fig.~\ref{fig:parameters_heatmap}. 
An animation of this figure is given as an online animation where the 3D shape of the plots can be visualised. We provide a version of these plots where the plasma parameter values are given in direct units in Figs.~\ref{fig:plasma_parameters} and \ref{fig:parameters_heatmap_direct}. 

The passage of the large-scale coronal wave can be clearly distinguished as a peak in the 3D profiles of all three plasma parameters $\rho$, $\bar{T}$, and EM (Fig.~\ref{fig:plasma_parameters}) at 22:21:07~UT, and as a sharp bright front in the heatmaps (Fig.~\ref{fig:parameters_heatmap}). The density (panel a) and the EM (panel c) decrease strongly after the wave has passed, with the decrease being stronger in the western segments of the ROI (red) and shallower in the eastern ones (blue). A few minutes later, around 22:25~UT, the density and EM increase again in the western segments, while it decreases or stays constant in the remaining of the ROI. The temperature evolution (panel b) is considerably less smooth. While the arrival of the wave can be clearly distinguished as a peak, also in the heatmaps (see Figs.~\ref{fig:parameters_heatmap} and \ref{fig:parameters_heatmap_direct}), the centre segments (yellow-green) remain at enhanced temperatures for at least 4~minutes. The western (red) segments also show a second enhancement at 22:25~UT, while the eastern (blue) segments show a decrease in temperature after the wave has passed and a gradual increase thereafter. 
Notably, the three plasma parameters show very distinct profiles for different segments of the ROI, which illustrates the high variability in the response of the coronal plasma.

\begin{figure}[t]
    \centering
    \resizebox{\hsize}{!}{\includegraphics{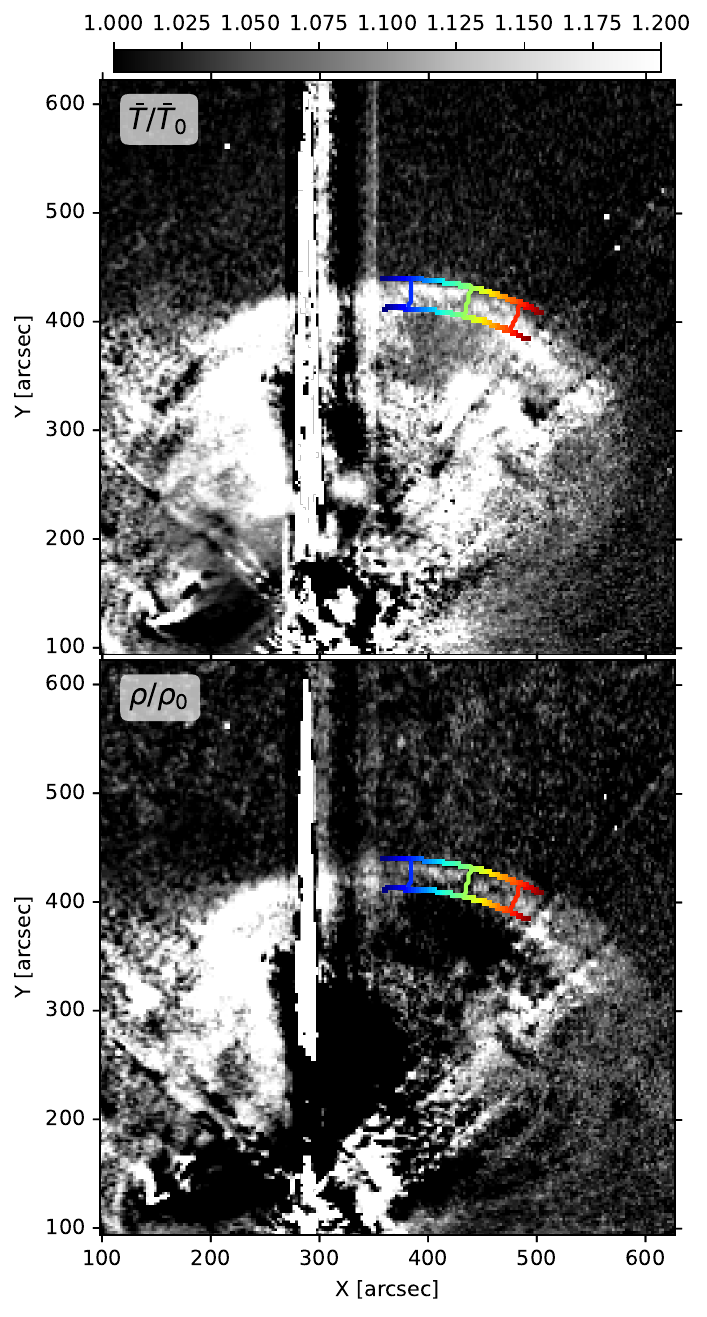}}
        \caption{Base-ratio maps of the mean weighted temperature $\bar{T}$ and density $\rho$ at 22:21:07~UT. The rainbow-colour contours outline the arc-shaped ROI in Fig.~\ref{fig:overview}. The associated movie is available online.}
        \label{fig:ratio_snapshot}
\end{figure}

To further investigate this variance in the plasma response to the large-scale coronal wave at different points in space, Fig.~\ref{fig:ratio_snapshot} shows the base-ratio temperature and density maps at 22:21:07~UT, i.e. when the wave passes over the ROI and causes the sharp increase in the plasma parameter evolution. The accompanying animation shows the time evolution of the maps during the analysis time frame. The two rainbow-coloured arcs mark the ROI region shown in Fig.~\ref{fig:overview}. As previously described, the plasma parameters increase the most in the western region and a dip can be clearly distinguished in the centre regions. The temperature increases around 13\% in the whole ROI, with maximum increase of 16\% and minimum of 8\%. The density shows the strongest variance, increasing between 10\% in some regions and none in others. The sharp gap in the increase of both the temperature and density can be clearly distinguished in the maps in the centre of the ROI, which is an indication of how small the scale of the variability of the coronal response can be.

\begin{figure*}[h!]
    \centering
    \resizebox{\hsize}{!}{\includegraphics{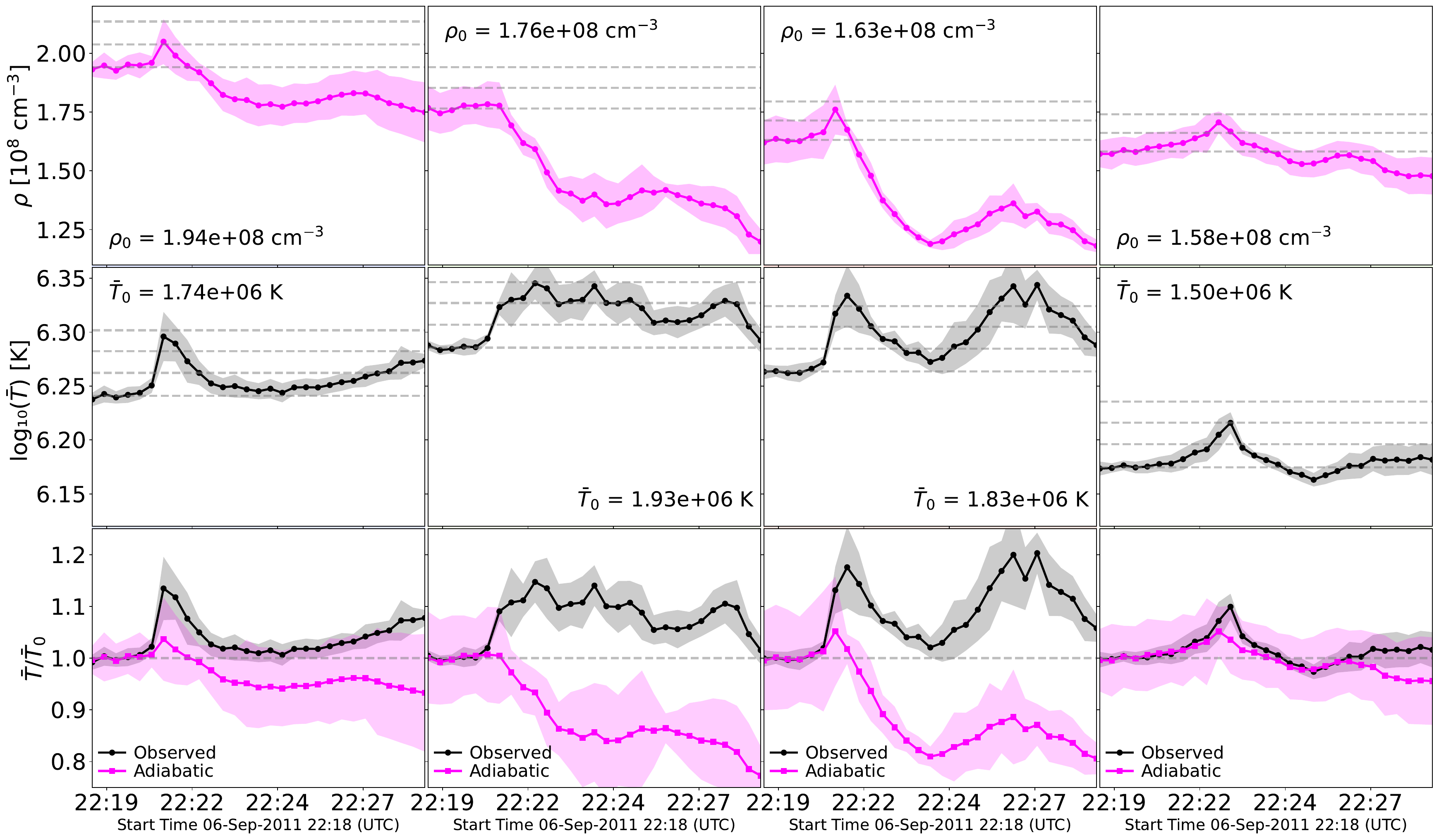}}
        \caption{Time evolution of the density $\rho$ (top row) and mean weighted temperature $\bar{T}$ (middle row) for four representative patches along the EUV wave propagation (indicated in Fig.~\ref{fig:overview}). Columns correspond to the east, centre, west, and north patches (from left to right, as defined in Fig.~\ref{fig:overview}). Horizontal dashed grey lines mark 5\% increments relative to the pre-event values. In the bottoms panel, the relative changes of the DEM-derived mean weighted temperature are plotted in black, while the relative changes of the temperature predicted from the adiabatic relation using the measured density evolution is overplotted in magenta. Peak values for each patch are listed in Table~\ref{tab:parameters}.}
        \label{fig:adiabatic_heating}
\end{figure*}

Figure~\ref{fig:adiabatic_heating} shows the evolution of the density and temperature for the selected east (first column), centre (second column), west (third column) and north (fourth column) patches. The horizontal dashed gray lines show relative increases of 5\% with respect to the pre-event values of each profile. The peak values of the plasma parameters in each patch are summarised in Table~\ref{tab:parameters}. In the east patch of the ROI (first column) the density increases from $1.94\times10^{8}$ to $2.05\times10^{8}$~cm$^{-3}$, which implies a 6\% increase; in the west patch (third column), the density goes from $1.63\times10^{8}$ to $1.76\times10^{8}$~cm$^{-3}$, which represents a 8\% increase; in the centre patch (second column) the density starts at $1.76\times10^{8}$~cm$^{-3}$ and increases on average barely 1\%, but decreases right after with the expansion of the coronal dimming. 

The temperature, however, increases considerably in all three patches of the ROI: the west patch increases from a temperature of $1.83\times10^6$ to $2.16\times10^6$~K, the centre patch from $1.93\times10^6$ to $2.21\times10^6$~K, and the east patch from $1.74\times10^6$ to $1.98\times10^6$~K. These values represent an increase of 18, 15, and 13\%, respectively. 
In the case of the west and east patches, the temperature decreases immediately after the large-scale coronal wave has passed reaching $1.87\times10^6$ and $1.76\times10^6$~K at 22:24:19~UT, respectively. It then gradually increases in the east patch, and does so strongly in the west patch up to $2.21\times10^6$~K. The centre patch remains at temperatures above $2.21\times10^6$~K for over 5~minutes after the wave has passed. This lasting high temperature in the centre parts of the ROI can also be clearly distinguished as a bright stripe in the temperature heatmaps shown in Figs.~\ref{fig:parameters_heatmap} and \ref{fig:parameters_heatmap_direct}. 

The fourth column in Fig.~\ref{fig:adiabatic_heating} shows the density and temperature evolution for the north patch, which is positioned further away from the AR. The initial values are notably lower than in the ROI, with $\rho=1.58\times10^8$~cm$^{-3}$ and $\bar{T}=1.50\times10^6$~K. In this patch, the wave produces a density increase of 8\% and a temperature increase of 10\% (see Table~\ref{tab:parameters}).  

\subsubsection{Adiabatic prediction}
Several studies have shown that the temperature increase in the front of a large-scale coronal wave can often be explained by adiabatic heating of the plasma caused by compression (\citet{schrijver2011february, downs2012understanding, vanninathan2015coronal}). This implies that, considering an initial temperature $\bar{T}_0$ and density $\rho_0$ of the plasma, the wave should increase the temperature following the adiabatic relation:
\begin{equation}\label{eq:adiabatic}
    \frac{T}{T_0}=\left(\frac{\rho}{\rho_0}\right)^{\gamma-1},
\end{equation}
where $\gamma$ is the adiabatic index for fully ionised plasma, $\gamma=5/3$. The bottom row in Fig.~\ref{fig:adiabatic_heating} shows the temperature of the four patches as extracted from the DEM analysis with the mean temperature $\bar{T}$ derived from the observations according to Eq.~(\ref{eq:t}) plotted as black line alongside the temperature predicted using the adiabatic equation (Eq.~\ref{eq:adiabatic}) in magenta. We calculated the initial temperature $\bar{T}_0$ and density $\rho_0$ as the average of the first five values of the time series, i.e. over the pre-event phase. 

In all four patches the adiabatic prediction underestimates the temperature increase caused by the large-scale coronal wave. In the edge patches of the ROI (first and third column) adiabatic heating could potentially account for the initial temperature increase, given the large uncertainties in the density, but it cannot explain the total temperature increase caused by the wave. The most pronounced difference appears in the centre patch, where the temperature increases strongly while the density shows only a slight increase or no increase at all. This indicates that compression alone cannot account for the observed temperature increase, and that additional processes such as magnetic reconnection, maybe initiated by the passing wave front, likely play a significant role. In the north patch, the adiabatic prediction is closest to the observations. In fact, in the initial temperature and density increase, the predicted temperature under adiabatic conditions closely aligns with the observed values, and only during the peak of the temperature it does not hold any more. We note that the peak of the temperature increase occurs after the wave has already passed the patch. 
In summary, this analysis shows that compression alone cannot explain the observed temperature increases, specially in the ROI, where the temperature increases are very large compared to the density increase. 

\subsection{DEM curve response to a large-scale coronal wave} \label{sec:res:dem}

\begin{figure*}[h!]
    \centering
    \resizebox{\hsize}{!}{\includegraphics{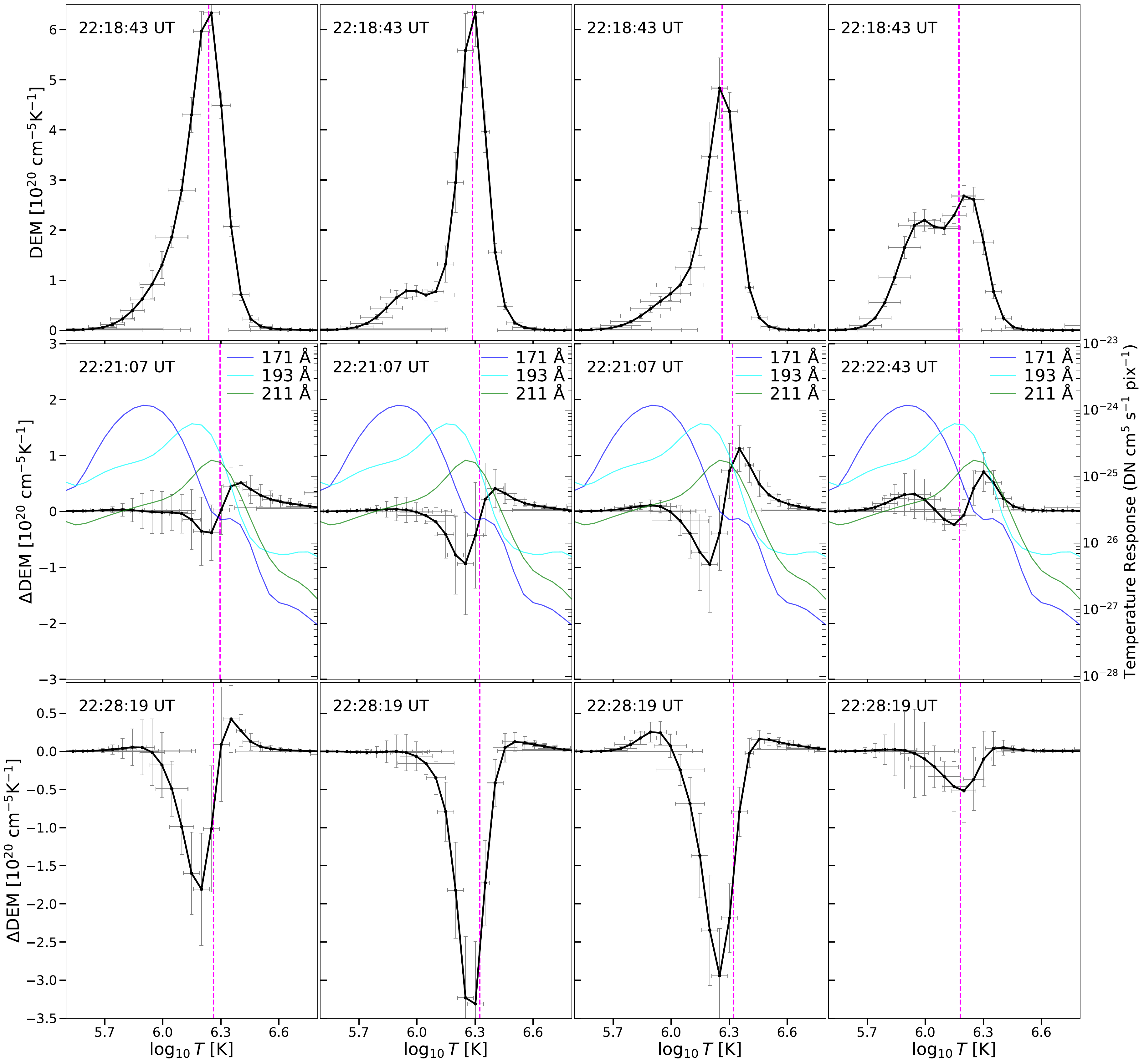}}
        \caption{DEM distributions for the four representative patches (columns: east, centre, west, and north; see Fig.~\ref{fig:overview}). The top row shows the DEM curves before the arrival of the wave (22:18:43~UT). The middle and bottom rows show the background-subtracted DEM ($\Delta$DEM), obtained by subtracting the pre-event DEM from the DEM at the time of the wave front arrival at 22:21:07~UT (22:22:43~UT for the north patch) and at a later time at 22:28:19~UT, respectively. Vertical magenta lines indicate the mean weighted temperature $\bar{T}$ for each patch and time. In the middle row, the $\Delta$DEM curves are plotted together with the AIA temperature response functions for the 171, 193, and 211~\AA~channels.}
        \label{fig:dem_curves}
\end{figure*}

Figure~\ref{fig:dem_curves} shows the DEM and $\Delta$DEM curves at different times in the four patches. The top row shows the DEM curve of the four patches at 22:18:43~UT; i.e. before the arrival of the wave. The middle and bottom row show the background-subtracted DEM, $\Delta$DEM, where the DEM at 22:18:43~UT has been subtracted from the DEM at 22:21:07~UT (22:22:43~UT for the north patch) and 22:28:19~UT, respectively. In each panel, the vertical magenta line shows the mean weighted temperature $\bar{T}$ in the respective patches.

The DEM curve response to the large-scale coronal wave show the heating of plasma from lower to higher temperatures, with amplitudes of $\Delta\text{DEM}$ strongly dependent on the patch. The heating signature is the strongest in the west patch, weaker in the east, and in the centre accompanied by a net decrease, indicating both heating and plasma depletion. At later times, all patches exhibit a broad negative $\Delta$DEM consistent with a coronal dimming, which is most pronounced inside the ROI and weakest in the north patch. The minimum and maximum values of $\Delta$DEM for each patch at the passage of the EUV wave are given in Table~\ref{tab:parameters}.

Before the arrival of the wave (22:18:43~UT) the east (first column) and west (third column) patches have a slightly skewed Gaussian DEM. The centre (second column) and north (fourth column) segment shows an asymmetric bimodal DEM, specially asymmetric in the case of the centre patch. The change in the DEM distribution ($\Delta$DEM) caused by the large-scale coronal wave passing over the patches is shown in the middle row. The four patches show an antisymmetric distribution expected from the compressive heating that the wave produces: plasma at lower temperatures heats up to higher temperatures. This heating is most prominent in the west patch, with plasma at around $\log_{10}T=6.20$~K being heated up to $\log_{10}T=6.35$~K and the $\Delta$DEM reaching values of $\pm1\times10^{20}$~cm$^{-5}$K$^{-1}$. The change is least prominent in the east patch where plasma at $\log_{10}T=6.25$~K heats up to $\log_{10}T=6.40$~K, but $\Delta$DEM reaches lower values of $\pm0.5\times10^{20}$~cm$^{-5}$K$^{-1}$. The centre patch (second column) also shows plasma being heated from $\log_{10}T=6.25$ to $\log_{10}T=6.40$~K. However, the decrease is stronger ($\Delta\text{DEM}=-0.9\times10^{20}$) than the increase ($\Delta\text{DEM}=0.4\times10^{20}$), which implies that, not only plasma has been heated to higher temperatures, but also some of the plasma at around $\log_{10}T=6.25$~K has been depleted. 

The $\Delta$DEM at 22:28:19~UT shows a predominant inverted gaussian curve in all patches, reaching values of $-2.94$, $-3.31$, $-1.81$, and $-0.51$~cm$^{-5}$K$^{-1}$ in the west, centre, east, and north patch, respectively. This shape of the curve is produced by the density depletion resulting from the expansion and depletion of plasma due to the CME, which is also visible as a coronal dimming in Fig.~\ref{fig:overview} and its accompanying animation. The dimming can be seen to be less pronounced in the north patch. 

Plasma heating to higher temperatures has been proposed as an explanation for why large-scale coronal waves sometimes appear as intensity decreases in the 171~\AA~channel \citep{schrijver2011february, downs2012understanding, vanninathan2015coronal}. As the plasma temperature rises, the peak of the DEM curve shifts toward temperature regions where the 171~\AA~channel response function is lower. The change in emission is proportional to the local gradient of the response function, therefore the compressed but also heated wave front can appear as a dark feature. The second row of Fig.~\ref{fig:dem_curves} shows the AIA response function for the 171, 193 and 211~\AA~channels together with the $\Delta$DEM. The initial temperatures (vertical magenta line) in all patches lie close to the plateau of the 171~\AA~response, beyond the peak of the 193~\AA~response, and close to the peak of the 211~\AA~response. This indicates that when the plasma is compressed and heated, the temperature increase occurs along the declining gradient of the response curve of the 193~\AA~channel, and along the plateau of 171~\AA~channel. Consequently, a decrease in 171~\AA~intensity is not expected, consistent with the observations by \citet{dissauer2016projection}. In contrast, the 193~\AA~channel may be affected, with the temperature shift partially offsetting the intensity enhancement produced by density compression. For the 211~\AA~channel, the initial mean weighted temperature $\bar{T}$ is located near the peak of the response function, and shifts only slightly down the gradient in all three ROI patches, most noticeably in the centre patch. 

The north patch $\Delta$DEM at 22:22:43~UT shown in the fourth column, gives a useful reference point to the changes observed in the ROI. In the north patch, the initial mean weighted temperature $\bar{T}$ is lower (see Sect.~\ref{sec:res:param} and Table~\ref{tab:parameters}) and in consequence lies below the peak in the response function of the 211~\AA~channel and in the declining gradient of the 171~\AA~channel. However, in this case the increase in $\bar{T}$ is not notable and thus we do not expect shifting in the response curve to have a notable effect. 

Apart from the shift of the mean weighted temperature along the response functions, we also considered the negative and positive $\Delta\text{DEM}$ peaks caused by the coronal wave heating the plasma. In the case of the west patch (third column), these two peaks cross the 211~\AA~response function at very similar values from $1.5\times10^{-25}$ to $1.0\times10^{-25}$~DN~cm$^5$~s$^{-1}$~pix$^{-1}$. In the other two ROI cases (centre and east patches) the second peak, which shows the temperature at which most of the plasma is added, occurs considerably lower in the gradient at $5.1\times10^{-26}$~DN~cm$^5$~s$^{-1}$~pix$^{-1}$. Similarly to the 171~\AA~decrease wave observations, plasma heating could also have caused a decrease in brightness enhancement in the centre and east patches, as the plasma was heated to temperatures with lower response in the 211~\AA~channel. 

This hypothesis agrees with observations, as the wave is barely visible as an enhancement in 171 and 193~\AA~observations in the ROI patches. Furthermore, the wave is more prominent in the west patch than in the east one in 211~\AA~observations, and is not visible in the centre patch. As previously discussed, the centre patch does not only shift towards temperatures with a lower response, but it also has a notable density depletion which is the main contributor to the localised temporary disappearance of the wave front.

\section{Summary and discussion} \label{sec:discussion}
We present a study on the coronal plasma response to the large-scale coronal wave on 6 September 2011, associated with an X2.1 class flare and CME event. We employed SDO/AIA observations in all six coronal EUV passbands and perform DEM reconstruction using the regularised inversion algorithm developed by \citet{hannah2012differential}. We derived DEM curves and plasma parameters ($\rho$, $\bar{T}$ and EM) over four narrow patches and an extended region along the wave front. This approach enabled us to quantify plasma changes in selected regions, and examine the spatial variability in the plasma parameters across the wave front, and its visibility.

We found that when the wave is clearly detected in the AIA 211 and 335~\AA~channels, the plasma density increases by 6-8\% (see Table~\ref{tab:parameters}). This range is consistent with previous studies of strong large-scale coronal waves, including \citet{vanninathan2015coronal} (6-9\%), \citet{veronig2011plasma} (~$\approx$10\%), \citet{schrijver2011february} ($\approx$10\%), and \citet{zhang2025responses} ($\approx$8\%, within a coronal hole). The values in this study lie at the lower end of results obtained from DEM analyses of large-scale coronal wave shock fronts above the solar limb identified through Type II radio bursts, such as \citet{kozarev2011offlimb} (12-18\%) and \citet{frassati2020estimate} (15-23\%). This difference is expected, as our analysis does not specifically target shock fronts, where larger density enhancements are expected.

We obtained temperature increases along the wave front of 10-18\% (Table~\ref{tab:parameters}), exceeding the values reported by \citet{schrijver2011february} ($\approx$7\%) and \citet{vanninathan2015coronal} (5-6\%), and comparable to those found by \citet{zhang2025responses} ($\approx$12\%). Across all patches, the observed temperature rise is greater than that predicted from adiabatic compression alone based on the measured density increase. Assuming purely adiabatic compression of the plasma, the expected temperature increase would be approximately $\approx$5\%, which considerably underestimates the observed values.

\citet{vanninathan2015coronal} examined two regions along different large-scale coronal wave propagation paths in the 15 September 2011 event, and evaluated whether the observed heating could be explained by compression. They found that the initial temperature increase closely resembled adiabatic predictions, but it deviated from the expectations in the relaxation phase after the wave had already surpassed the region. A similar trend was reported by \citet{zhang2025responses} for the response of coronal hole plasma to an EUV wave, where initially the temperature increased adiabatically, but additional increases were observed later.
For the 6 September 2011 large-scale coronal wave, our results indicate that adiabatic heating accounts only for the initial stage of the temperature rise, while it fails to predict the peak and the relaxation phase of the temperature evolution (see Fig.~\ref{fig:adiabatic_heating}). This behaviour varies across patches: in the initially hotter patches ($\bar{T}>1.8$~MK, west and centre patches) the discrepancy with the adiabatic prediction is larger, whereas in cooler patches ($\bar{T}<1.75$~MK, east and north patches) the agreement is better. In the north patch, the adiabatic estimate closely matches the measurements except at the temperature peak.

The excess heating may arise from non-linear effects not captured by Eq.~\ref{eq:adiabatic}, as well as from uncertainties in the density computation. In particular, the density enhancement at the wave front may be underestimated due to the overlap of the wave pulse with the coronal dimming developing at higher altitudes. Furthermore, the density estimates rely on the assumption of a constant LOS thickness of the perturbation ($h\approx170$~Mm). If the actual thickness of the wave front is smaller, as suggested by \citet{dissauer2016projection}, the corresponding density enhancement would be larger (see Eq.~\ref{eq:n}), and therefore the adiabatic prediction would fit better the observed temperature increase.

Additional heating may also result from local energy release triggered by the wave front. For example, the large-scale coronal wave may disturb local magnetic field lines and reconnect with favourably oriented ones \citep{attrill2007coronal} increasing the local temperature without affecting the density. Another possible explanation is wave mode conversion of the incoming wave. This process can occur when a fast-mode wave encounters regions of weak magnetic field, where the Alfvén speed is comparable to the sound speed, such as near magnetic null points and quasi-separatrix layers (QSLs). Under these conditions, part of the incoming fast-mode wave may convert into slow-mode waves that subsequently propagate along, or become trapped in, the local magnetic field \citep{chen2016can, chandra2018observations, kumar2024direct}. 
Specifically, this interpretation could explain the prolonged temperature enhancement observed in the central sectors of the ROI, as evidenced by Figs.~\ref{fig:plasma_parameters}, \ref{fig:parameters_heatmap}, and \ref{fig:parameters_heatmap_direct}. In these regions, the temperature remains at a $\approx$10\% enhancement for 5~minutes after the wave passage, while the density decreases by approximately 30\%, which would not be possible without alternative heating sources in addition to the compressional heating by the wave.

\begin{figure}[t]
    \centering
    \resizebox{\hsize}{!}{\includegraphics{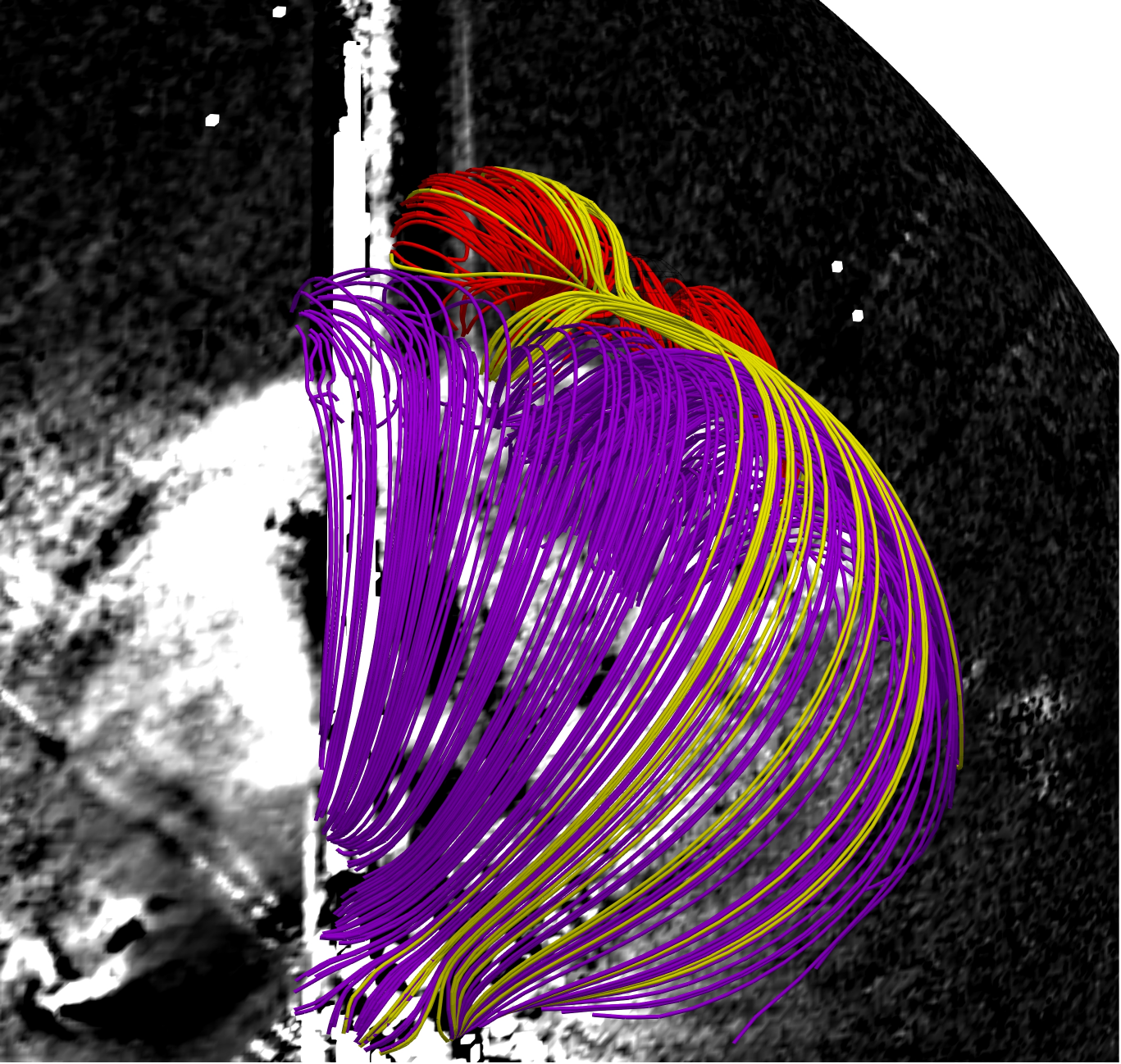}}
        \caption{PFSS extrapolation of field lines surrounding the ROI over the temperature base-ratio map shown in Fig.~\ref{fig:ratio_snapshot}. Purple and red field lines show separate magnetic flux systems, while the yellow field lines show field lines going around a magnetic null point.}
        \label{fig:pfss_extrapolation}
\end{figure}

To test whether this could be the case, Fig.~\ref{fig:pfss_extrapolation} shows the potential field source-surface (PFSS) extrapolation around the AR, computed using the \texttt{pfsspy} Python package \citep{Stansby2020pfsspy}. The background image displays the temperature base-ratio map from Fig.~\ref{fig:ratio_snapshot}, showing the wave front at 22:21~UT. The purple field lines originate from the AR and extend northwards, with the end points landing roughly ahead of the wave front, while red field lines correspond to lower lying loops located north of the front and expanding further northwards. This configuration suggests the presence of a QSL between the two flux systems. Moreover, a magnetic null point appears to be located close to the centre of the wave front, near the yellow field lines, thus indicating that the central part of the ROI is favourable for mode conversion. This interpretation is further supported by the faint, slowly propagating intensity enhancement visible in Fig.~\ref{fig:aia_overview}b ahead of the dimming region, which may correspond to a newly formed slow-mode wave.

A similar mechanism may also contribute to the second enhancement observed in the western segments (see red lines in Fig.~\ref{fig:plasma_parameters}). As seen in the animation accompanying Fig.~\ref{fig:overview}, the fast-mode wave propagating northwestwards impacts coronal loops, after which a secondary, conjectured slow-mode wave is generated and appears to propagate eastwards along the loops towards the western segments, which could cause the temperature and density enhancement observed. However, the animation accompanying Fig.~\ref{fig:app:overview} also shows hot plasma being ejected northwards from the AR, which could also cause such an enhancement.

The central ROI segments present an additional observational peculiarity: the wave temporarily `disappears' in the 171, 193 and 211~\AA~channels, while it remains visible in the 335~\AA~channel. \citet{dissauer2016projection} studied this event and, after ruling out interaction with other magnetic structures, attributed this observations to LOS effects. In the 6 September 2011 event, a coronal dimming closely follows the expansion of the large-scale coronal wave \citep{dissauer2018detection, prasad2020mhd}. Using quadrature observations from SDO and STEREO-A, \citet{dissauer2016projection} found that the reduced intensity (observed as coronal dimmings) from the mass depletion associated with the accompanying CME, and the increased intensity from the large-scale coronal wave could add up to background intensity, effectively resulting in the loss of the wave signature. Since coronal emission is optically thin,  the observed intensity would correspond to the LOS integration of the coronal dimming occurring at greater heights in the corona and the wave propagating at lower heights. Because coronal dimmings are most prominent at quiet-Sun temperatures (e.g. \citet{dissauer2018statistics}), this effect would not have a strong impact in channels sensitive to hotter plasma, such as 335~\AA. However, at the precise times of the wave disappearance no STEREO-A data is available to completely exclude additional effects.

A potential contributor for a decrease in the wave front visibility could be plasma heating at the wave front that shifts the emission away from the peak of the response function of a passband. We considered this possibility by analysing the response of the DEM curve to the large-scale coronal wave, shown in Fig.~\ref{fig:dem_curves}. In the centre of the ROI, the initial mean weighted temperature ($\bar{T}_0=1.93$~MK) lies beyond the peak of the AIA 193~\AA~response function, around the peak of 211~\AA, and in the plateau of 171~\AA. Additional heating at the wave front would therefore move the emission down the 211~\AA~response gradient, reducing the signal in the 211~\AA~observations. However, the plasma temperature in the west and east patches of the ROI is also notably high, particularly in the west patch ($\bar{T}_0=1.83$~MK), but the wave front remains visible. This suggests that, while plasma heating might have taken a role in the disappearance observations, it was not the main contributor. Nonetheless, such effects may explain the weak signatures observed in the 171 and 193~\AA~channels (see \citet{dissauer2016projection}). In addition, recently \citet{li2025dependence} reported on the intensity dependency of large-scale coronal waves on the local magnetic field inclination. Although this possibility lies beyond the scope of the present work, it may also contribute to the decrease (and reappearance) of the wave signal.

We also analysed the spatial variability of the plasma parameters across the wave front within the ROI (see Figs.~\ref{fig:plasma_parameters} and \ref{fig:ratio_snapshot}). The central region -- where the wave disappears -- shows the smallest density increase when the wave passes over the region. The density variation over the ROI shown in Fig.~\ref{fig:ratio_snapshot} exhibits a pronounced dip around the centre segments (green and yellow), with increases below 1\%. We further interpret this variability in the context of the magnetic field configuration shown in Fig.~\ref{fig:pfss_extrapolation}. As discussed previously, a null point is located near the centre of the ROI. Owing to this topology, some of the yellow field lines in Fig.~\ref{fig:pfss_extrapolation} extend westward above the western segments, whereas the eastern field lines connect back to the AR, similar to the purple field lines. Such magnetic configuration may naturally produce spatial variability in the plasma response in several ways: different magnetic flux systems participate in the dimming differently, leading to variations in the intensity decrease across the ROI; additionally, wave mode conversion is expected to be more efficient in the vicinity of the null point, while wave-induced magnetic reconnection will only occur in favourably oriented field lines, both of these processes could cause variations in the plasma response.
These observations highlight the strong spatial variability in the plasma response along the wave front, even over relatively small angular extents. This variability is important to consider, as many large-scale coronal wave studies rely on integrations over large, potentially non-coherent, regions.

Finally, we note that the DEM analysis of fast large-scale coronal waves has an inherent methodological limitation. DEM reconstruction requires combining images from multiple wavelength channels that are not strictly cotemporal, while the wave front propagates between exposures. As a result, the wave appears at different locations in different channels, which limits the accuracy of the DEM reconstruction. This limitation could only be overcome with fully simultaneous multiwavelength observations. Consequently, we also needed to integrate along the wave path to capture the wave front across all wavelengths.

\section{Conclusion} \label{sec:conclusions}
In this work, we present the analysis of the coronal plasma response to the passage of the large-scale coronal wave associated with the X2.1 class flare and CME on 6 September 2011. This event had the particularity of the wave front temporarily disappearing in some SDO/AIA channels (211, 193~\AA) while staying visible in other channels (like 335~\AA). We employed DEM analysis to extract the plasma parameters ($\rho$, $\bar{T}$, and EM) at a variety of locations, such as different regions along the wave front when it disappears and further out in its propagation. We also analysed how the DEM response curve changes with the wave compressing the plasma by its passage. Our main findings are:
\begin{itemize}
    \item The large-scale coronal wave produces enhancements of 6-8\% in density, of 10-18\% in temperature, and of 12-16\% in emission measure.  
    \item Adiabatic heating caused by plasma compression can only explain the initial temperature increase, but not the peak nor relaxation phase of the temperature evolution, which indicates that alternative heating mechanisms must have played a role in the temperature increase. The deviation from the adiabatic prediction is stronger in regions with higher initial plasma temperature. 
    \item Background-subtracted DEM curves ($\Delta\text{DEM}$) show plasma heating from lower to higher temperatures at the wave front, and later times show a broad-band negative $\Delta\text{DEM}$ consistent with a coronal dimming.
    \item Temporal evolution of the density and emission measure show the mass depletion associated with the initiation of the CME-produced coronal dimming above the selected locations. The temperature shows stronger variability, with some sectors remaining at an increased temperature for 5~minutes, which point to wave mode conversion at a QSL.
    \item In the region of the wave front disappearance, the plasma parameter changes caused by the wave show strong spatial variability. For instance, in some regions the density shows no increase at all and 10\% in others. The variability is also notable at very fine scales.
    \item The DEM curve shows that the initial temperature in the affected region is notably higher than quiet-Sun regions ($\bar{T}\gtrsim1.7$~MK), which allows plasma to be heated beyond the peak response of the AIA 193 and 211~\AA~channels. This effect might have contributed in the temporary decrease and disappearance of the wave signal in those channels.
\end{itemize}

Overall, this study provides quantitative measurements of the coronal response to a large-scale coronal wave and a spatially resolved analysis of plasma properties at the wave front. The results support a scenario in which both wave-driven plasma compression and CME-associated coronal dimming contribute to the observed signature, consistent with the findings by \citet{dissauer2016projection}. They further demonstrate that the coronal response can vary significantly on small spatial scales, highlighting the importance of local plasma conditions and magnetic field configuration in the interpretation of large-scale coronal wave observations.

\begin{acknowledgements}
This project has received funding from the European Union's Horizon Europe research and innovation programme under grant agreement No 101134999 (SOLER). The research was sponsored by the DynaSun project and has thus received funding under the Horizon Europe programme of the European Union under grant agreement (no. 101131534). Views and opinions expressed are however those of the author(s) only and do not necessarily reflect those of the European Union and therefore the European Union cannot be held responsible for them. SDO data are courtesy of NASA/SDO and the AIA. We are grateful to Dr. Mariana Cécere for the insightful discussions on wave propagation in different magnetic environments. We also thank the referee for all the suggestions and comments, which have constructively helped improve and strengthen the manuscript.
\end{acknowledgements}

\bibliographystyle{aa}
\bibliography{My_References}

\begin{thebibliography}{96}
\expandafter\ifx\csname natexlab\endcsname\relax\def\natexlab#1{#1}\fi

\bibitem[{{Asai} {et~al.}(2008){Asai}, {Hara}, {Watanabe}, {Imada}, {Sakao}, {Narukage}, {Culhane}, \& {Doschek}}]{asai2008strongly}
{Asai}, A., {Hara}, H., {Watanabe}, T., {et~al.} 2008, \apj, 685, 622

\bibitem[{{Asai} {et~al.}(2012){Asai}, {Ishii}, {Isobe}, {Kitai}, {Ichimoto}, {UeNo}, {Nagata}, {Morita}, {Nishida}, {Shiota}, {Oi}, {Akioka}, \& {Shibata}}]{asai2012first}
{Asai}, A., {Ishii}, T.~T., {Isobe}, H., {et~al.} 2012, \apjl, 745, L18

\bibitem[{{Aschwanden}(2005)}]{aschwanden2005physics}
{Aschwanden}, M.~J. 2005, {Physics of the Solar Corona. An Introduction with Problems and Solutions} (Springer)

\bibitem[{{Attrill} {et~al.}(2007){Attrill}, {Harra}, {van Driel-Gesztelyi}, \& {D{\'e}moulin}}]{attrill2007coronal}
{Attrill}, G. D.~R., {Harra}, L.~K., {van Driel-Gesztelyi}, L., \& {D{\'e}moulin}, P. 2007, \apjl, 656, L101

\bibitem[{{Boerner} {et~al.}(2012){Boerner}, {Edwards}, {Lemen}, {Rausch}, {Schrijver}, {Shine}, {Shing}, {Stern}, {Tarbell}, {Title}, {Wolfson}, {Soufli}, {Spiller}, {Gullikson}, {McKenzie}, {Windt}, {Golub}, {Podgorski}, {Testa}, \& {Weber}}]{boerner2012initial}
{Boerner}, P., {Edwards}, C., {Lemen}, J., {et~al.} 2012, \solphys, 275, 41

\bibitem[{{Chandra} {et~al.}(2018){Chandra}, {Chen}, {Joshi}, {Joshi}, \& {Schmieder}}]{chandra2018observations}
{Chandra}, R., {Chen}, P.~F., {Joshi}, R., {Joshi}, B., \& {Schmieder}, B. 2018, \apj, 863, 101

\bibitem[{{Chen}(2016)}]{chen2016global}
{Chen}, P.~F. 2016, Geophysical Monograph Series, 216, 381

\bibitem[{{Chen} {et~al.}(2016){Chen}, {Fang}, {Chandra}, \& {Srivastava}}]{chen2016can}
{Chen}, P.~F., {Fang}, C., {Chandra}, R., \& {Srivastava}, A.~K. 2016, \solphys, 291, 3195

\bibitem[{{Chen} {et~al.}(2002){Chen}, {Wu}, {Shibata}, \& {Fang}}]{chen2002evidence}
{Chen}, P.~F., {Wu}, S.~T., {Shibata}, K., \& {Fang}, C. 2002, \apjl, 572, L99

\bibitem[{{Chen} \& {Wu}(2011)}]{chen2011first}
{Chen}, P.~F. \& {Wu}, Y. 2011, \apjl, 732, L20

\bibitem[{{Cheng} {et~al.}(2012{\natexlab{a}}){Cheng}, {Zhang}, {Olmedo}, {Vourlidas}, {Ding}, \& {Liu}}]{cheng2012investigation}
{Cheng}, X., {Zhang}, J., {Olmedo}, O., {et~al.} 2012{\natexlab{a}}, \apjl, 745, L5

\bibitem[{{Cheng} {et~al.}(2012{\natexlab{b}}){Cheng}, {Zhang}, {Saar}, \& {Ding}}]{cheng2012differential}
{Cheng}, X., {Zhang}, J., {Saar}, S.~H., \& {Ding}, M.~D. 2012{\natexlab{b}}, \apj, 761, 62

\bibitem[{{Del Zanna} {et~al.}(2021){Del Zanna}, {Dere}, {Young}, \& {Landi}}]{zanna2021chianti}
{Del Zanna}, G., {Dere}, K.~P., {Young}, P.~R., \& {Landi}, E. 2021, \apj, 909, 38

\bibitem[{{Delaboudini{\`e}re} {et~al.}(1995){Delaboudini{\`e}re}, {Artzner}, {Brunaud}, {Gabriel}, {Hochedez}, {Millier}, {Song}, {Au}, {Dere}, {Howard}, {Kreplin}, {Michels}, {Moses}, {Defise}, {Jamar}, {Rochus}, {Chauvineau}, {Marioge}, {Catura}, {Lemen}, {Shing}, {Stern}, {Gurman}, {Neupert}, {Maucherat}, {Clette}, {Cugnon}, \& {Van Dessel}}]{delaboudiniere1995eit}
{Delaboudini{\`e}re}, J.-P., {Artzner}, G.~E., {Brunaud}, J., {et~al.} 1995, \solphys, 162, 291

\bibitem[{{Delann{\'e}e} {et~al.}(2014){Delann{\'e}e}, {Artzner}, {Schmieder}, \& {Parenti}}]{delannee2014time}
{Delann{\'e}e}, C., {Artzner}, G., {Schmieder}, B., \& {Parenti}, S. 2014, \solphys, 289, 2565

\bibitem[{{Delann{\'e}e} \& {Aulanier}(1999)}]{delanee1999cme}
{Delann{\'e}e}, C. \& {Aulanier}, G. 1999, \solphys, 190, 107

\bibitem[{{Dennis} \& {Phillips}(2024)}]{brian2024emission}
{Dennis}, B.~R. \& {Phillips}, K. J.~H. 2024, \solphys, 299, 48

\bibitem[{{Dere} {et~al.}(1997){Dere}, {Landi}, {Mason}, {Monsignori Fossi}, \& {Young}}]{dere1997chianti}
{Dere}, K.~P., {Landi}, E., {Mason}, H.~E., {Monsignori Fossi}, B.~C., \& {Young}, P.~R. 1997, \aaps, 125, 149

\bibitem[{{Dissauer} {et~al.}(2016){Dissauer}, {Temmer}, {Veronig}, {Vanninathan}, \& {Magdaleni{\'c}}}]{dissauer2016projection}
{Dissauer}, K., {Temmer}, M., {Veronig}, A.~M., {Vanninathan}, K., \& {Magdaleni{\'c}}, J. 2016, \apj, 830, 92

\bibitem[{{Dissauer} {et~al.}(2018{\natexlab{a}}){Dissauer}, {Veronig}, {Temmer}, {Podladchikova}, \& {Vanninathan}}]{dissauer2018detection}
{Dissauer}, K., {Veronig}, A.~M., {Temmer}, M., {Podladchikova}, T., \& {Vanninathan}, K. 2018{\natexlab{a}}, \apj, 855, 137

\bibitem[{{Dissauer} {et~al.}(2018{\natexlab{b}}){Dissauer}, {Veronig}, {Temmer}, {Podladchikova}, \& {Vanninathan}}]{dissauer2018statistics}
{Dissauer}, K., {Veronig}, A.~M., {Temmer}, M., {Podladchikova}, T., \& {Vanninathan}, K. 2018{\natexlab{b}}, \apj, 863, 169

\bibitem[{{Domingo} {et~al.}(1995){Domingo}, {Fleck}, \& {Poland}}]{domingo1995soho}
{Domingo}, V., {Fleck}, B., \& {Poland}, A.~I. 1995, \solphys, 162, 1

\bibitem[{{Downs} {et~al.}(2012){Downs}, {Roussev}, {van der Holst}, {Lugaz}, \& {Sokolov}}]{downs2012understanding}
{Downs}, C., {Roussev}, I.~I., {van der Holst}, B., {Lugaz}, N., \& {Sokolov}, I.~V. 2012, \apj, 750, 134

\bibitem[{{Eto} {et~al.}(2002){Eto}, {Isobe}, {Narukage}, {Asai}, {Morimoto}, {Thompson}, {Yashiro}, {Wang}, {Kitai}, {Kurokawa}, \& {Shibata}}]{eto2002relation}
{Eto}, S., {Isobe}, H., {Narukage}, N., {et~al.} 2002, \pasj, 54, 481

\bibitem[{{Feng} {et~al.}(2013){Feng}, {Wiegelmann}, {Su}, {Inhester}, {Li}, {Sun}, \& {Gan}}]{feng2013magnetic}
{Feng}, L., {Wiegelmann}, T., {Su}, Y., {et~al.} 2013, \apj, 765, 37

\bibitem[{{Frassati} {et~al.}(2020){Frassati}, {Mancuso}, \& {Bemporad}}]{frassati2020estimate}
{Frassati}, F., {Mancuso}, S., \& {Bemporad}, A. 2020, \solphys, 295, 124

\bibitem[{{Gallagher} \& {Long}(2011)}]{gallagher2011large}
{Gallagher}, P.~T. \& {Long}, D.~M. 2011, \ssr, 158, 365

\bibitem[{{Gopalswamy} {et~al.}(2009){Gopalswamy}, {Yashiro}, {Temmer}, {Davila}, {Thompson}, {Jones}, {McAteer}, {Wuelser}, {Freeland}, \& {Howard}}]{gopalswamy2009euv}
{Gopalswamy}, N., {Yashiro}, S., {Temmer}, M., {et~al.} 2009, \apjl, 691, L123

\bibitem[{{Hannah} \& {Kontar}(2012)}]{hannah2012differential}
{Hannah}, I.~G. \& {Kontar}, E.~P. 2012, \aap, 539, A146

\bibitem[{{Harra} {et~al.}(2011){Harra}, {Sterling}, {G{\"o}m{\"o}ry}, \& {Veronig}}]{harra2011spectroscopic}
{Harra}, L.~K., {Sterling}, A.~C., {G{\"o}m{\"o}ry}, P., \& {Veronig}, A. 2011, \apjl, 737, L4

\bibitem[{{Hudson} {et~al.}(1996){Hudson}, {Acton}, \& {Freeland}}]{hudson1996long}
{Hudson}, H.~S., {Acton}, L.~W., \& {Freeland}, S.~L. 1996, \apj, 470, 629

\bibitem[{{Janvier} {et~al.}(2016){Janvier}, {Savcheva}, {Pariat}, {Tassev}, {Millholland}, {Bommier}, {McCauley}, {McKillop}, \& {Dougan}}]{janvier2016evolution}
{Janvier}, M., {Savcheva}, A., {Pariat}, E., {et~al.} 2016, \aap, 591, A141

\bibitem[{{Kienreich} {et~al.}(2009){Kienreich}, {Temmer}, \& {Veronig}}]{keinreich2009stereo}
{Kienreich}, I.~W., {Temmer}, M., \& {Veronig}, A.~M. 2009, \apjl, 703, L118

\bibitem[{{Klassen} {et~al.}(2000){Klassen}, {Aurass}, {Mann}, \& {Thompson}}]{klassen2000catalogue}
{Klassen}, A., {Aurass}, H., {Mann}, G., \& {Thompson}, B.~J. 2000, \aaps, 141, 357

\bibitem[{{Kozarev} {et~al.}(2011){Kozarev}, {Korreck}, {Lobzin}, {Weber}, \& {Schwadron}}]{kozarev2011offlimb}
{Kozarev}, K.~A., {Korreck}, K.~E., {Lobzin}, V.~V., {Weber}, M.~A., \& {Schwadron}, N.~A. 2011, \apjl, 733, L25

\bibitem[{{Kumar} {et~al.}(2013){Kumar}, {Cho}, {Chen}, {Bong}, \& {Park}}]{kumar2013multiwavelength}
{Kumar}, P., {Cho}, K.-S., {Chen}, P.~F., {Bong}, S.-C., \& {Park}, S.-H. 2013, \solphys, 282, 523

\bibitem[{{Kumar} \& {Manoharan}(2013)}]{kumar2013eruption}
{Kumar}, P. \& {Manoharan}, P.~K. 2013, \aap, 553, A109

\bibitem[{{Kumar} {et~al.}(2024){Kumar}, {Nakariakov}, {Karpen}, \& {Cho}}]{kumar2024direct}
{Kumar}, P., {Nakariakov}, V.~M., {Karpen}, J.~T., \& {Cho}, K.-S. 2024, Nature Communications, 15, 2667

\bibitem[{{Lemen} {et~al.}(2012){Lemen}, {Title}, {Akin}, {Boerner}, {Chou}, {Drake}, {Duncan}, {Edwards}, {Friedlaender}, {Heyman}, {Hurlburt}, {Katz}, {Kushner}, {Levay}, {Lindgren}, {Mathur}, {McFeaters}, {Mitchell}, {Rehse}, {Schrijver}, {Springer}, {Stern}, {Tarbell}, {Wuelser}, {Wolfson}, {Yanari}, {Bookbinder}, {Cheimets}, {Caldwell}, {Deluca}, {Gates}, {Golub}, {Park}, {Podgorski}, {Bush}, {Scherrer}, {Gummin}, {Smith}, {Auker}, {Jerram}, {Pool}, {Soufli}, {Windt}, {Beardsley}, {Clapp}, {Lang}, \& {Waltham}}]{Lemen2012aia}
{Lemen}, J.~R., {Title}, A.~M., {Akin}, D.~J., {et~al.} 2012, \solphys, 275, 17

\bibitem[{Li {et~al.}(2025)Li, Zheng, Wang, \& Chen}]{li2025observations}
Li, J., Zheng, R., Wang, X., \& Chen, Y. 2025, Science China Physics, Mechanics \& Astronomy, 68, 259611

\bibitem[{{Li} {et~al.}(2025){Li}, {Guo}, {Ni}, {Zhang}, \& {Chen}}]{li2025dependence}
{Li}, Y.~W., {Guo}, J.~H., {Ni}, Y.~W., {Zhang}, Z.~Y., \& {Chen}, P.~F. 2025, \aap, 698, A316

\bibitem[{{Liu} {et~al.}(2019){Liu}, {Wang}, {Lee}, \& {Shen}}]{liu2019impacts}
{Liu}, R., {Wang}, Y., {Lee}, J., \& {Shen}, C. 2019, \apj, 870, 15

\bibitem[{{Liu} \& {Ofman}(2014)}]{liu2014advances}
{Liu}, W. \& {Ofman}, L. 2014, \solphys, 289, 3233

\bibitem[{{Liu} {et~al.}(2012){Liu}, {Ofman}, {Nitta}, {Aschwanden}, {Schrijver}, {Title}, \& {Tarbell}}]{liu2012quasi}
{Liu}, W., {Ofman}, L., {Nitta}, N.~V., {et~al.} 2012, \apj, 753, 52

\bibitem[{{Long} {et~al.}(2017){Long}, {Bloomfield}, {Chen}, {Downs}, {Gallagher}, {Kwon}, {Vanninathan}, {Veronig}, {Vourlidas}, {Vr{\v{s}}nak}, {Warmuth}, \& {{\v{Z}}ic}}]{long2017understanding}
{Long}, D.~M., {Bloomfield}, D.~S., {Chen}, P.~F., {et~al.} 2017, \solphys, 292, 7

\bibitem[{{Long} {et~al.}(2011){Long}, {DeLuca}, \& {Gallagher}}]{long2011wave}
{Long}, D.~M., {DeLuca}, E.~E., \& {Gallagher}, P.~T. 2011, \apjl, 741, L21

\bibitem[{{Ma} {et~al.}(2011){Ma}, {Raymond}, {Golub}, {Lin}, {Chen}, {Grigis}, {Testa}, \& {Long}}]{ma2011observations}
{Ma}, S., {Raymond}, J.~C., {Golub}, L., {et~al.} 2011, \apj, 738, 160

\bibitem[{{Mann} {et~al.}(1999){Mann}, {Aurass}, {Klassen}, {Estel}, \& {Thompson}}]{mann1999coronal}
{Mann}, G., {Aurass}, H., {Klassen}, A., {Estel}, C., \& {Thompson}, B.~J. 1999, in ESA Special Publication, Vol. 446, 8th SOHO Workshop: Plasma Dynamics and Diagnostics in the Solar Transition Region and Corona, ed. J.~C. {Vial} \& B.~{Kaldeich-Sch{\"u}}, 477

\bibitem[{{Mann} \& {Veronig}(2023)}]{mann2023propagation}
{Mann}, G. \& {Veronig}, A.~M. 2023, \aap, 676, A144

\bibitem[{{Moreton} \& {Ramsey}(1960)}]{moreton1960recent}
{Moreton}, G.~E. \& {Ramsey}, H.~E. 1960, \pasp, 72, 357

\bibitem[{{Moses} {et~al.}(1997){Moses}, {Clette}, {Delaboudini{\`e}re}, {Artzner}, {Bougnet}, {Brunaud}, {Carabetian}, {Gabriel}, {Hochedez}, {Millier}, {Song}, {Au}, {Dere}, {Howard}, {Kreplin}, {Michels}, {Defise}, {Jamar}, {Rochus}, {Chauvineau}, {Marioge}, {Catura}, {Lemen}, {Shing}, {Stern}, {Gurman}, {Neupert}, {Newmark}, {Thompson}, {Maucherat}, {Portier-Fozzani}, {Berghmans}, {Cugnon}, {Van Dessel}, \& {Gabryl}}]{moses1997eit}
{Moses}, D., {Clette}, F., {Delaboudini{\`e}re}, J.-P., {et~al.} 1997, \solphys, 175, 571

\bibitem[{{Muhr} {et~al.}(2011){Muhr}, {Veronig}, {Kienreich}, {Temmer}, \& {Vr{\v{s}}nak}}]{muhr2011analysis}
{Muhr}, N., {Veronig}, A.~M., {Kienreich}, I.~W., {Temmer}, M., \& {Vr{\v{s}}nak}, B. 2011, \apj, 739, 89

\bibitem[{{Muhr} {et~al.}(2014){Muhr}, {Veronig}, {Kienreich}, {Vr{\v{s}}nak}, {Temmer}, \& {Bein}}]{muhr2014statistical}
{Muhr}, N., {Veronig}, A.~M., {Kienreich}, I.~W., {et~al.} 2014, \solphys, 289, 4563

\bibitem[{{Nitta} {et~al.}(2014){Nitta}, {Liu}, {Gopalswamy}, \& {Yashiro}}]{nitta2014relation}
{Nitta}, N.~V., {Liu}, W., {Gopalswamy}, N., \& {Yashiro}, S. 2014, \solphys, 289, 4589

\bibitem[{{Nitta} {et~al.}(2013){Nitta}, {Schrijver}, {Title}, \& {Liu}}]{nitta2013large}
{Nitta}, N.~V., {Schrijver}, C.~J., {Title}, A.~M., \& {Liu}, W. 2013, \apj, 776, 58

\bibitem[{{Okamoto} {et~al.}(2004){Okamoto}, {Nakai}, {Keiyama}, {Narukage}, {UeNo}, {Kitai}, {Kurokawa}, \& {Shibata}}]{okamoto2004filament}
{Okamoto}, T.~J., {Nakai}, H., {Keiyama}, A., {et~al.} 2004, \apj, 608, 1124

\bibitem[{{Olmedo} {et~al.}(2012){Olmedo}, {Vourlidas}, {Zhang}, \& {Cheng}}]{olmedo2012secondary}
{Olmedo}, O., {Vourlidas}, A., {Zhang}, J., \& {Cheng}, X. 2012, \apj, 756, 143

\bibitem[{{Patsourakos} \& {Vourlidas}(2012)}]{patsourakos2012nature}
{Patsourakos}, S. \& {Vourlidas}, A. 2012, \solphys, 281, 187

\bibitem[{{Patsourakos} {et~al.}(2009){Patsourakos}, {Vourlidas}, {Wang}, {Stenborg}, \& {Thernisien}}]{patsourakos2009what}
{Patsourakos}, S., {Vourlidas}, A., {Wang}, Y.~M., {Stenborg}, G., \& {Thernisien}, A. 2009, \solphys, 259, 49

\bibitem[{{Pesnell} {et~al.}(2012){Pesnell}, {Thompson}, \& {Chamberlin}}]{pesnell2012sdo}
{Pesnell}, W.~D., {Thompson}, B.~J., \& {Chamberlin}, P.~C. 2012, \solphys, 275, 3

\bibitem[{{Podladchikova} \& {Berghmans}(2005)}]{podladchikova2005automated}
{Podladchikova}, O. \& {Berghmans}, D. 2005, \solphys, 228, 265

\bibitem[{{Podladchikova} {et~al.}(2019){Podladchikova}, {Veronig}, {Dissauer}, {Temmer}, \& {Podladchikova}}]{podladchikova2019three}
{Podladchikova}, T., {Veronig}, A.~M., {Dissauer}, K., {Temmer}, M., \& {Podladchikova}, O. 2019, \apj, 877, 68

\bibitem[{{Prasad} {et~al.}(2020){Prasad}, {Dissauer}, {Hu}, {Bhattacharyya}, {Veronig}, {Kumar}, \& {Joshi}}]{prasad2020mhd}
{Prasad}, A., {Dissauer}, K., {Hu}, Q., {et~al.} 2020, \apj, 903, 129

\bibitem[{{Purkhart} {et~al.}(2023){Purkhart}, {Veronig}, {Dickson}, {Battaglia}, {Krucker}, {Jarolim}, {Kliem}, {Dissauer}, {Podladchikova}, {STIX Team}, \& {EUI Team}}]{purkhart2023multipoint}
{Purkhart}, S., {Veronig}, A.~M., {Dickson}, E. C.~M., {et~al.} 2023, \aap, 679, A99

\bibitem[{{Qu} {et~al.}(2017){Qu}, {Jiang}, \& {Chen}}]{qu2017observations}
{Qu}, Z.~N., {Jiang}, L.~Q., \& {Chen}, S.~L. 2017, \apj, 851, 41

\bibitem[{{Razquin} {et~al.}(2026){Razquin}, {Dissauer}, {Veronig}, \& {Barnes}}]{arazquin2026selected}
{Razquin}, A., {Dissauer}, K., {Veronig}, A.~M., \& {Barnes}, G. 2026, \aap, 708, A292

\bibitem[{{Schrijver} {et~al.}(2011){Schrijver}, {Aulanier}, {Title}, {Pariat}, \& {Delann{\'e}e}}]{schrijver2011february}
{Schrijver}, C.~J., {Aulanier}, G., {Title}, A.~M., {Pariat}, E., \& {Delann{\'e}e}, C. 2011, \apj, 738, 167

\bibitem[{{Schrijver} {et~al.}(2013){Schrijver}, {Title}, {Yeates}, \& {DeRosa}}]{schrijver2013pathways}
{Schrijver}, C.~J., {Title}, A.~M., {Yeates}, A.~R., \& {DeRosa}, M.~L. 2013, \apj, 773, 93

\bibitem[{{Shen} {et~al.}(2014){Shen}, {Ichimoto}, {Ishii}, {Tian}, {Zhao}, \& {Shibata}}]{shen2014chain}
{Shen}, Y., {Ichimoto}, K., {Ishii}, T.~T., {et~al.} 2014, \apj, 786, 151

\bibitem[{{Shen} \& {Liu}(2012)}]{shen2012evidence}
{Shen}, Y. \& {Liu}, Y. 2012, \apj, 754, 7

\bibitem[{{Shen} {et~al.}(2013){Shen}, {Liu}, {Su}, {Li}, {Zhao}, {Tian}, {Ichimoto}, \& {Shibata}}]{shen2013diffraction}
{Shen}, Y., {Liu}, Y., {Su}, J., {et~al.} 2013, \apjl, 773, L33

\bibitem[{Stansby {et~al.}(2020)Stansby, Yeates, \& Badman}]{Stansby2020pfsspy}
Stansby, D., Yeates, A., \& Badman, S.~T. 2020, Journal of Open Source Software, 5, 2732

\bibitem[{{Sterling} \& {Hudson}(1997)}]{sterling1997yohkoh}
{Sterling}, A.~C. \& {Hudson}, H.~S. 1997, \apjl, 491, L55

\bibitem[{{Thompson} {et~al.}(1999){Thompson}, {Gurman}, {Neupert}, {Newmark}, {Delaboudini{\`e}re}, {Cyr}, {Stezelberger}, {Dere}, {Howard}, \& {Michels}}]{thompson1999soho}
{Thompson}, B.~J., {Gurman}, J.~B., {Neupert}, W.~M., {et~al.} 1999, \apjl, 517, L151

\bibitem[{{Thompson} {et~al.}(1998){Thompson}, {Plunkett}, {Gurman}, {Newmark}, {St. Cyr}, \& {Michels}}]{thompson1998soho}
{Thompson}, B.~J., {Plunkett}, S.~P., {Gurman}, J.~B., {et~al.} 1998, \grl, 25, 2465

\bibitem[{{Tripathi} \& {Raouafi}(2007)}]{tripathi2007on}
{Tripathi}, D. \& {Raouafi}, N.-E. 2007, \aap, 473, 951

\bibitem[{{Uchida}(1968)}]{uchida1968propagating}
{Uchida}, Y. 1968, \solphys, 4, 30

\bibitem[{{Vanninathan} {et~al.}(2015){Vanninathan}, {Veronig}, {Dissauer}, {Madjarska}, {Hannah}, \& {Kontar}}]{vanninathan2015coronal}
{Vanninathan}, K., {Veronig}, A.~M., {Dissauer}, K., {et~al.} 2015, \apj, 812, 173

\bibitem[{{Vanninathan} {et~al.}(2018){Vanninathan}, {Veronig}, {Dissauer}, \& {Temmer}}]{vanninathan2018plasma}
{Vanninathan}, K., {Veronig}, A.~M., {Dissauer}, K., \& {Temmer}, M. 2018, \apj, 857, 62

\bibitem[{{Veronig} {et~al.}(2025){Veronig}, {Dissauer}, {Kliem}, {Downs}, {Hudson}, {Jin}, {Osten}, {Podladchikova}, {Prasad}, {Qiu}, {Thompson}, {Tian}, \& {Vourlidas}}]{veronig2025coronal}
{Veronig}, A.~M., {Dissauer}, K., {Kliem}, B., {et~al.} 2025, Living Reviews in Solar Physics, 22, 2

\bibitem[{{Veronig} {et~al.}(2011){Veronig}, {G{\"o}m{\"o}ry}, {Kienreich}, {Muhr}, {Vr{\v{s}}nak}, {Temmer}, \& {Warren}}]{veronig2011plasma}
{Veronig}, A.~M., {G{\"o}m{\"o}ry}, P., {Kienreich}, I.~W., {et~al.} 2011, \apjl, 743, L10

\bibitem[{{Veronig} {et~al.}(2008){Veronig}, {Temmer}, \& {Vr{\v{s}}nak}}]{veronig2008high}
{Veronig}, A.~M., {Temmer}, M., \& {Vr{\v{s}}nak}, B. 2008, \apjl, 681, L113

\bibitem[{{Wang}(2000)}]{wang2000eit}
{Wang}, Y.-M. 2000, \apjl, 543, L89

\bibitem[{{Warmuth}(2007)}]{warmuth2007large}
{Warmuth}, A. 2007, in Lecture Notes in Physics, Berlin Springer Verlag, ed. K.-L. {Klein} \& A.~L. {MacKinnon}, Vol. 725, 107

\bibitem[{{Warmuth}(2015)}]{warmuth2015large}
{Warmuth}, A. 2015, Living Reviews in Solar Physics, 12, 3

\bibitem[{{Warmuth} \& {Mann}(2011)}]{warmuth2011kimnematical}
{Warmuth}, A. \& {Mann}, G. 2011, \aap, 532, A151

\bibitem[{{Warmuth} {et~al.}(2001){Warmuth}, {Vr{\v{s}}nak}, {Aurass}, \& {Hanslmeier}}]{warmuth2001evolution}
{Warmuth}, A., {Vr{\v{s}}nak}, B., {Aurass}, H., \& {Hanslmeier}, A. 2001, \apjl, 560, L105

\bibitem[{{Wills-Davey} \& {Attrill}(2009)}]{willsdavey2009eit}
{Wills-Davey}, M.~J. \& {Attrill}, G.~D.~R. 2009, \ssr, 149, 325

\bibitem[{{Wills-Davey} \& {Thompson}(1999)}]{willsdavey1999observations}
{Wills-Davey}, M.~J. \& {Thompson}, B.~J. 1999, \solphys, 190, 467

\bibitem[{{Wu} {et~al.}(2001){Wu}, {Zheng}, {Wang}, {Thompson}, {Plunkett}, {Zhao}, \& {Dryer}}]{wu2001three}
{Wu}, S.~T., {Zheng}, H., {Wang}, S., {et~al.} 2001, \jgr, 106, 25089

\bibitem[{{Zhang} {et~al.}(2025){Zhang}, {Chen}, {Zhou}, {Feng}, {Su}, {Guo}, {Li}, {Lin}, {Ma}, {Shen}, {Zheng}, {Liu}, {Bai}, {Deng}, \& {Wang}}]{zhang2025responses}
{Zhang}, X., {Chen}, H., {Zhou}, G., {et~al.} 2025, \apjl, 987, L3

\bibitem[{{Zhang} {et~al.}(2024){Zhang}, {Zhang}, {Song}, \& {Ji}}]{zhang2024transverse}
{Zhang}, Y., {Zhang}, Q., {Song}, D.-c., \& {Ji}, H. 2024, \apj, 963, 140

\bibitem[{{Zheng}(2024)}]{zheng2024recent}
{Zheng}, R. 2024, Proceedings of the Royal Society of London Series A, 480, 20230950

\bibitem[{{Zhou} {et~al.}(2020){Zhou}, {Gao}, {Wang}, {Lin}, {Su}, {Jin}, \& {Zhang}}]{zhou2020magnetic}
{Zhou}, G., {Gao}, G., {Wang}, J., {et~al.} 2020, \apj, 905, 150

\bibitem[{{Zhukov}(2011)}]{zhukov2011eit}
{Zhukov}, A.~N. 2011, Journal of Atmospheric and Solar-Terrestrial Physics, 73, 1096

\bibitem[{{Zhukov} \& {Auch{\`e}re}(2004)}]{zhukov2004nature}
{Zhukov}, A.~N. \& {Auch{\`e}re}, F. 2004, \aap, 427, 705

\end{thebibliography}

\begin{appendix}
\onecolumn
\section{Overview of the September 6 2011 wave in EUV observations}

\begin{figure}[h!]
    \centering
    \resizebox{0.8\hsize}{!}{\includegraphics{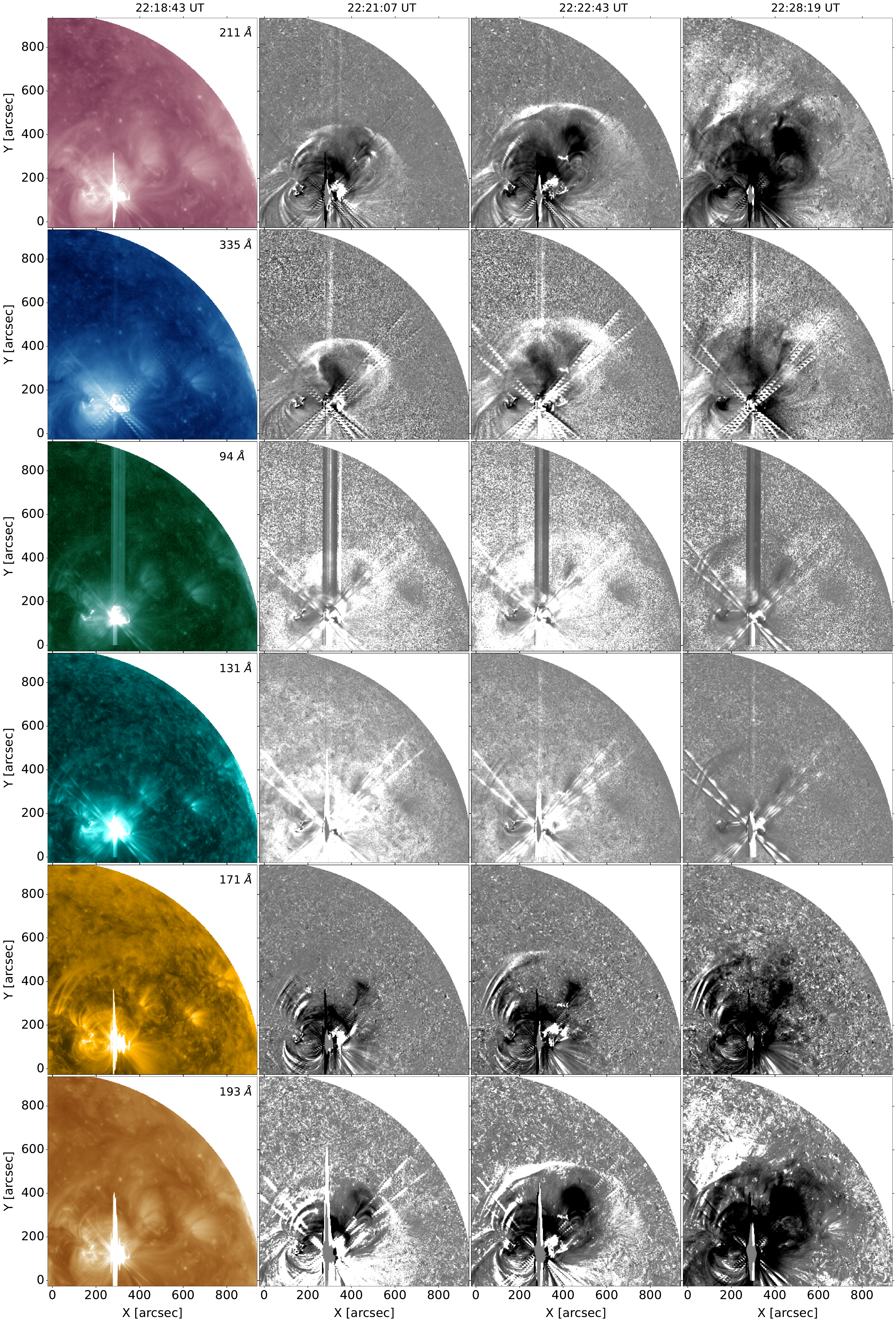}}
        \caption{Overview of the EUV wave event on 6 September 2011 observed with all EUV wavelengths of SDO/AIA in base-ratio images. The associated movie is available online}.
        \label{fig:app:overview}
\end{figure}
Figure~\ref{fig:app:overview} presents an extended version of Fig.~\ref{fig:overview} showing the large-scale coronal wave evolution on 6 September 2011 in all AIA EUV channels employed in the DEM reconstruction. The accompanying animation further illustrates the temporal evolution of the wave across the different channels. The wave is visible in all channels, and most predominantly in 193, 211, and 335~\AA. At 22:21:07~UT the wave front is absent in the 171, 193, and 211~\AA~channels. After the wave has already passed the ROI, some brightenings and plasma ejection can be observed north-west from the AR, which are the likely cause of the temperature increase observed in the western segments of the ROI (see Fig.~\ref{fig:plasma_parameters}).

\begin{figure*}[h]
    \centering
    \resizebox{\hsize}{!}{\includegraphics{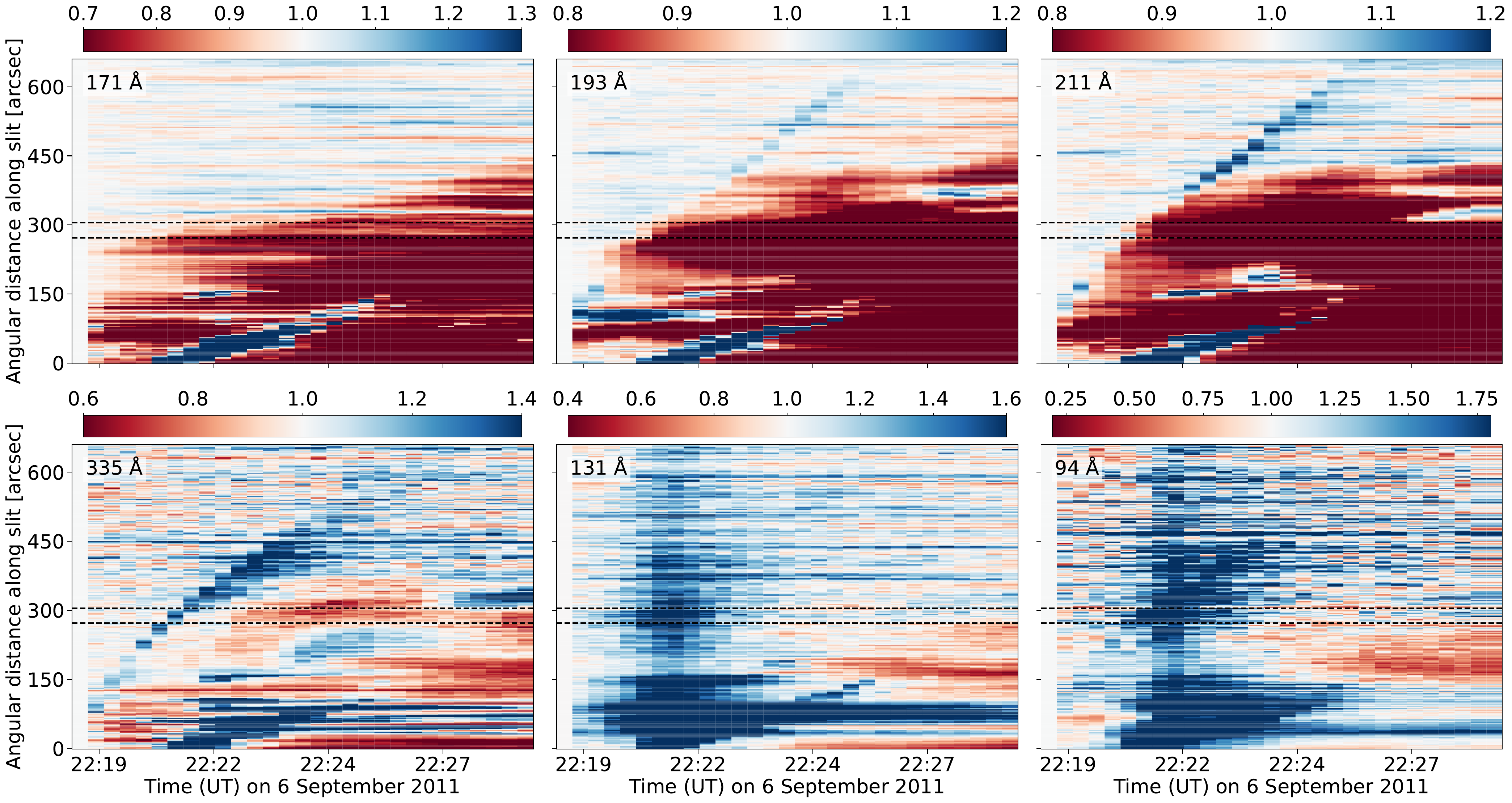}}
        \caption{Distance-time plots of the AIA EUV channel intensities along the slit indicated by the yellow line in Fig.~\ref{fig:overview}. Two horizontal black lines mark the portion of the slit corresponding to the ROI.}
        \label{fig:app:aia_overview}
\end{figure*}

Figure~\ref{fig:app:aia_overview} presents the intensity along the slit indicated in yellow in Fig.~\ref{fig:overview} for all AIA EUV channels. The propagating wave is clearly visible as an intensity enhancement (shown in blue) in the AIA 193, 211, and 335~\AA~channels, and is also distinguishable in the 94~\AA~channel. A coronal dimming, shown in red, dominates the evolution in the cooler channels (171, 193, and 211~\AA). This dimming appears to occur earlier in the 171~\AA~channel, while in the 193 and 211~\AA~channels it develops in part of the slit where the wave signal temporarily disappears. In the 94 and 131~\AA~channels, a generalised brightening of the slit is observed between 22:20 and 22:23~UT, which is associated with further flaring activity in the active region.

\section{Evolution of the plasma parameters}

\begin{figure}[t]
    \centering
    \resizebox{\hsize}{!}{\includegraphics{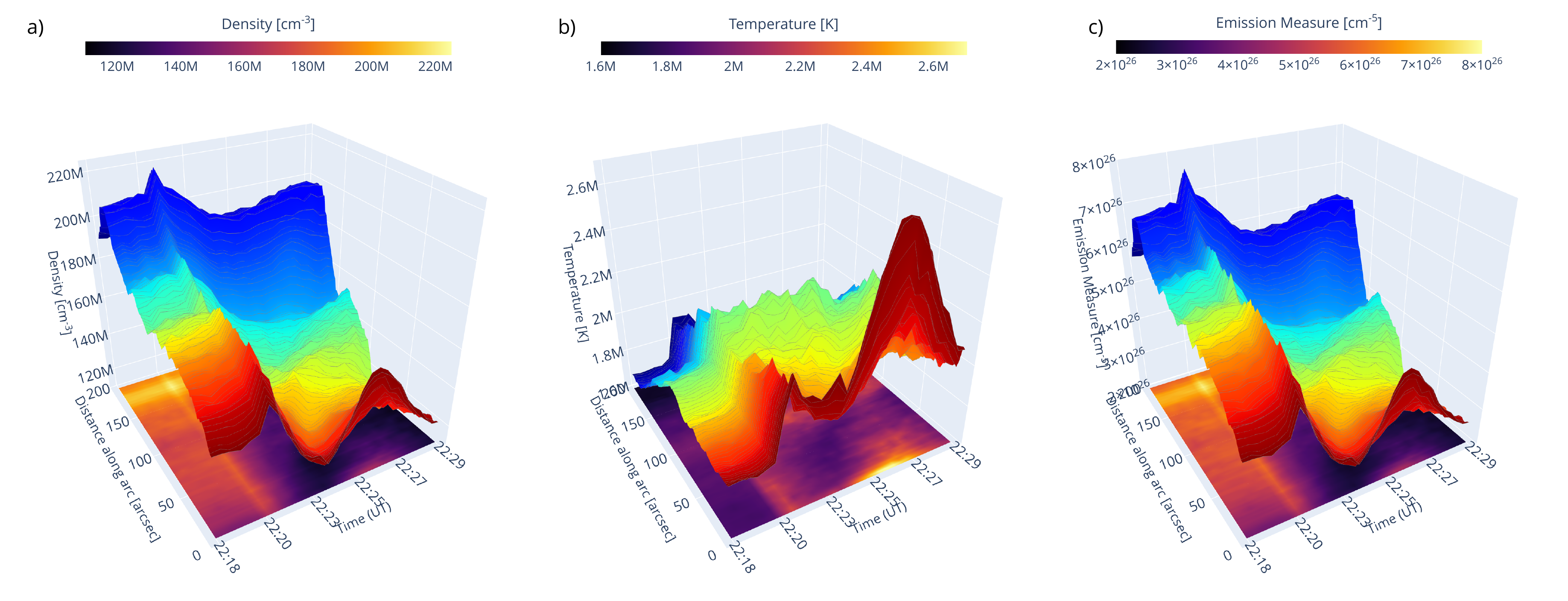}}
        \caption{Evolution of the plasma parameters derived from the DEM analysis within the arc-shaped ROI. Panels show the temporal variation of (a) density $\rho$, (b) mean weighted temperature $\bar{T}$, and (c) emission measure EM for 100 segments parallel to the wave propagation direction along the ROI. Individual curves are colour-coded by position along the arc, from west (red) to east (blue), consistent with the contour colour scale in Fig.~\ref{fig:overview}. The lower portion of each panel displays a distance–time heatmap of the corresponding relative changes. The associated movie is available online}.
        \label{fig:plasma_parameters_direct}
\end{figure}

The plasma parameters ($\rho$, $\bar{T}$, and EM) increase considerably with the passage of the large-scale coronal wave, as shown in Fig.~\ref{fig:plasma_parameters} and described in Sect.~\ref{sec:res:param}. Figure~\ref{fig:plasma_parameters_direct} shows the temporal evolution of the density $\rho$, mean emission-weighted temperature $\bar{T}$, and emission measure EM for all segments along the arc-shaped ROI in direct units. The increase in all plasma parameters caused by the wave is evident, regardless of the initial conditions. 

\begin{figure}[h]
    \centering
    \resizebox{0.8\hsize}{!}{\includegraphics{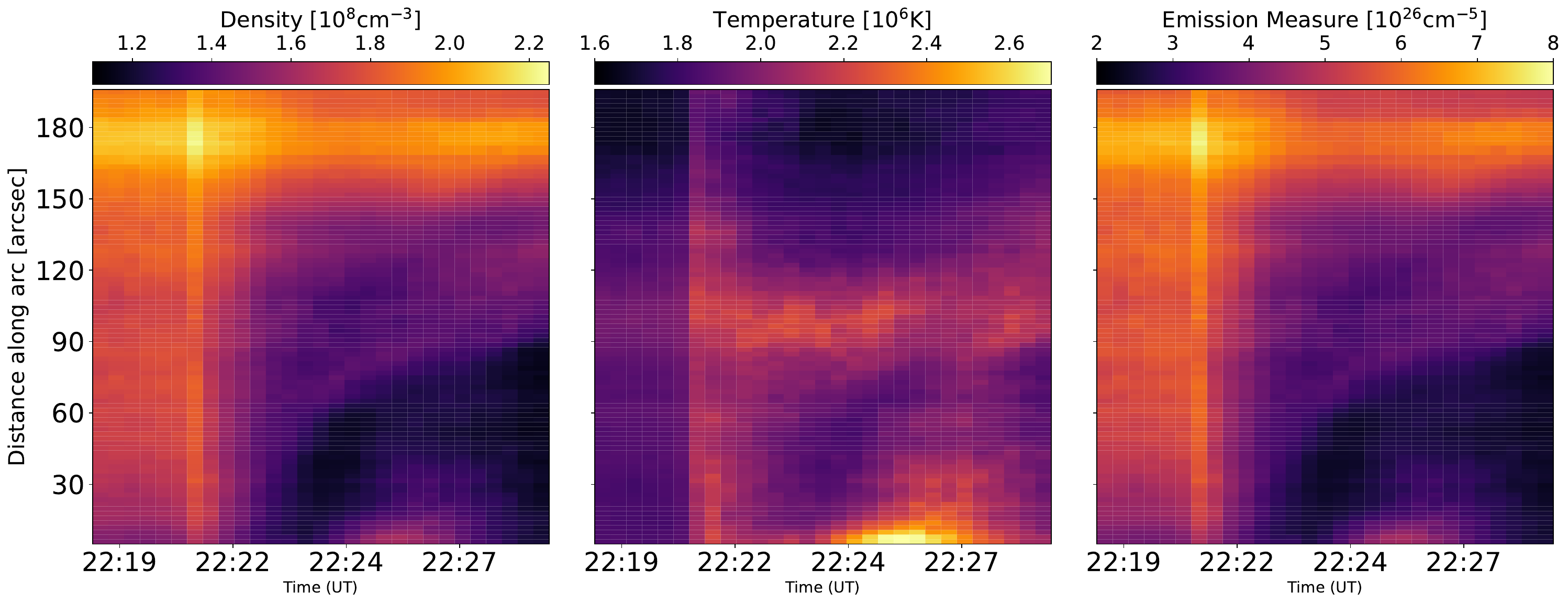}}
        \caption{Distance–time heatmaps of the evolution of the DEM-derived plasma parameters within the arc-shaped ROI. Panels show the temporal variation of (a) density $\rho$, (b) mean weighted temperature $\bar{T}$, and (c) emission measure EM for 100 segments oriented parallel to the wave propagation direction along the ROI. The vertical axis represents position along the arc from west to east.}
        \label{fig:parameters_heatmap_direct}
\end{figure}

Figure~\ref{fig:parameters_heatmap_direct} shows the corresponding distance–time heatmaps at the bottom boundary of Fig.~\ref{fig:plasma_parameters_direct}, revealing the propagation of the wave front as a coherent enhancement in all three parameters. The arrival of the wave to the ROI is visible in the three parameters at 22:21:07~UT, and it is specially notable for the temperature evolution. Figures~\ref{fig:plasma_parameters_direct} and \ref{fig:parameters_heatmap_direct} also illustrate that the density and emission measure are initially larger in the eastern segments of the ROI and decrease gradually towards the west, whereas the temperature is largest in the centre segments. 

\begin{figure}[h]
    \centering
    \resizebox{0.8\hsize}{!}{\includegraphics{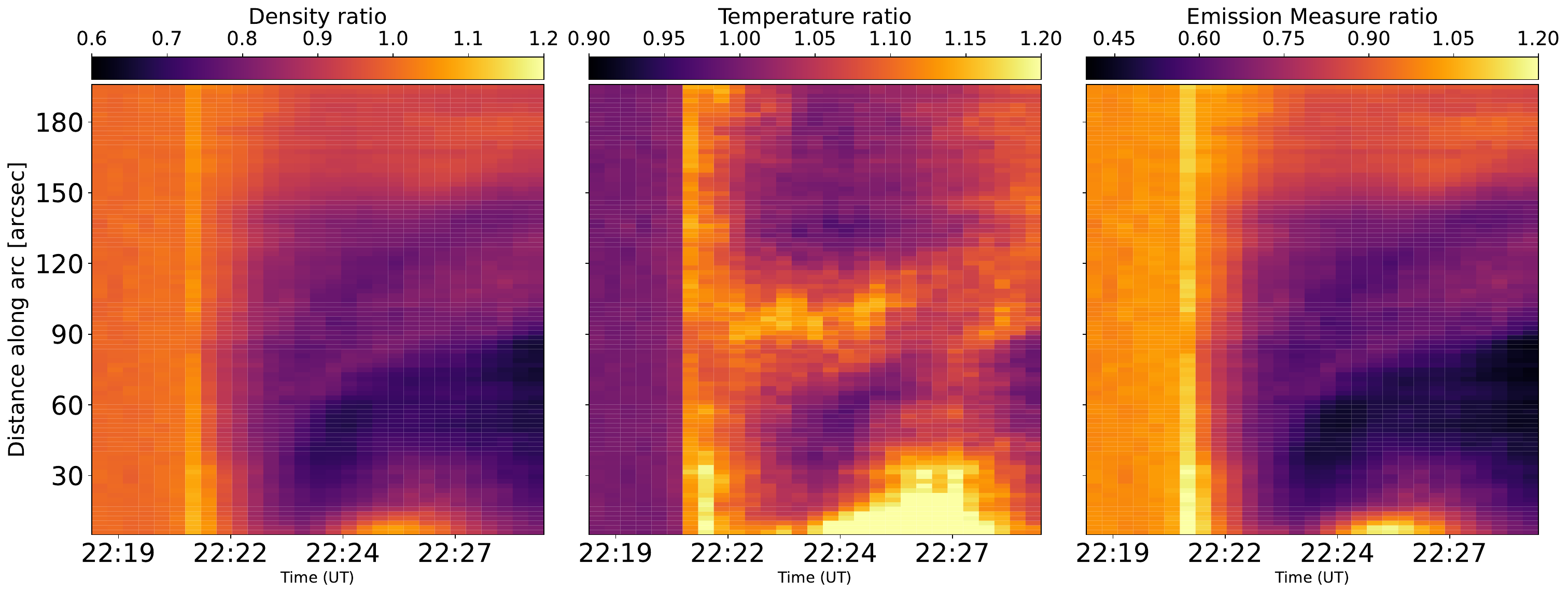}}
        \caption{Same as Fig.~\ref{fig:parameters_heatmap_direct} but showing the relative change of the plasma parameters ($\rho$, $\bar{T}$, and EM) with respect to the mean of the initial five values.}
        \label{fig:parameters_heatmap}
\end{figure}

Figure~\ref{fig:parameters_heatmap} shows the relative (initial values subtracted) distance-time heatmaps for the $\rho$, $\bar{T}$, and EM, which are also shown at the bottom boundary of the panels in Fig.~\ref{fig:plasma_parameters}. The increase is notable in the three plasma parameters, being stronger in the western segments of the ROI, and showing a dip in the centre segments. Both Fig.~\ref{fig:plasma_parameters_direct} and \ref{fig:plasma_parameters} show the lasting temperature increase in the centre segments as a bright yellow stripe.

\twocolumn
    
\end{appendix}

\end{document}